\documentclass[journal]{new-aiaa}
\usepackage[utf8]{inputenc}
\usepackage{textcomp}

\usepackage{graphicx}
\usepackage{amsmath}
\usepackage[version=4]{mhchem}
\usepackage{siunitx}
\usepackage{longtable,tabularx}
\usepackage{bm}
\usepackage{subcaption}
\usepackage{appendix}
\usepackage{algorithm}
\usepackage{algorithmic}

\usepackage{etoolbox}

\title{Star-based Navigation in the Outer Solar System}

\author{Vittorio Franzese \footnote{Research Scientist, Interdisciplinary Centre for Security, Reliability and Trust, University of Luxembourg, 29 Av. John F. Kennedy, email: vittorio.franzese@uni.lu. Corresponding author.}}
\affil{University of Luxembourg, 1855 Kirchberg, Luxembourg}

\begin{document}

\begin{center}
{\small
Accepted for publication in the \textit{Journal of Guidance, Control, and Dynamics}. This is the author's accepted manuscript.\\[-0.1cm]
The final version of record will be published by AIAA and will be available at the JGCD website.\\[-0.1cm]
Copyright © 2026 by the American Institute of Aeronautics and Astronautics, Inc. All rights reserved.
}
\end{center}

\vspace{-0.6em}

\maketitle

\begin{abstract}
This paper investigates an autonomous navigation method for spacecraft operating in the outer solar system, up to 250 AU from the Sun, using the parallactic shifts of nearby stars. These measurements enable estimation of the spacecraft trajectory while distant stars provide attitude information through conventional star-pattern matching. Stellar observation models are developed, accounting for delta light-time, parallax, and aberration effects. Navigation performance is assessed using two approaches: (1) a least-squares estimator using simultaneous multi-star measurements, and (2) a Kalman filter processing sequential single-star observations along deep-space trajectories. Monte Carlo simulations on trajectories representative of Voyager 1, Voyager 2, Pioneer 10, Pioneer 11, and New Horizons missions show sub-AU position accuracies at 250 AU, and velocity accuracies better than $4\cdot10^{-5}$ AU/day, under realistic spacecraft and instrumentation uncertainties. These values correspond to relative errors below 0.4\% in position and velocity with respect to the reference trajectories. Although less precise than radiometric tracking, this performance can support navigation in the outer solar system without reliance on Earth. When ground-based navigation remains necessary, this approach can be employed during long cruising phases, lowering the number of ground contacts. The method additionally shows potential for future missions venturing farther from the Sun.
\end{abstract}

\section*{Nomenclature}

{\renewcommand\arraystretch{1.0}
\begin{longtable*}{@{}l @{\quad=\quad} l@{}}

$A/m$ & area-to-mass ratio [m$^2$/kg] \\
$c$ & speed of light [km/s] \\
$c_r$ & reflectivity coefficient [-] \\
$d$ & mean Earth–Sun distance (1 AU) [AU] \\
$f$ & focal length [pixels] \\
$m$ & mass [kg] \\
$p$ & parallax [mas] \\
$r$ & heliocentric distance of spacecraft [AU] \\
$r_i$ & distance from Sun to $i$-th star [AU] \\
$t$ & time [s] \\
$t_0$ & catalog reference time [s] \\
$u,v$ & pixel coordinates [pixels] \\
$u_0, v_0$ & principal point coordinates [pixels] \\
$v_r$ & radial velocity [km/s] \\
$x_c, y_c, z_c$ & normalized camera coordinates [dimensionless] \\[3pt]

$S_0$ & solar constant at 1 AU [W/m$^2$] \\

$\bm{a}$ & acceleration vector [AU/day$^2$] \\
$\bm{a}_g$ & gravitational acceleration vector [AU/day$^2$] \\
$\bm{a}_s$ & solar radiation pressure acceleration vector [AU/day$^2$] \\
$\bm{f}$ & dynamics function vector \\
$\bm{h}$ & measurement function vector \\
$\bm{r}$ & spacecraft heliocentric position vector [AU] \\
$\bm{r}_i$ & heliocentric position vector of $i$-th star [AU] \\
$\bm{v}$ & spacecraft heliocentric velocity vector [AU/day] \\
$\bm{v}_i$ & heliocentric velocity vector of $i$-th star [AU/day] \\
$\bm{x}$ & state vector [AU, AU/day] \\
$\bm{y}$ & measurement vector \\[3pt]

$\bm{F}$ & dynamics Jacobian matrix \\
$\bm{F}_r$ & partial derivative of acceleration w.r.t. position \\
$\bm{H}$ & measurement Jacobian matrix \\
$\bm{I}$ & identity matrix \\
$\bm{K}$ & Kalman gain matrix \\
$\bm{L}_i$ & projection matrix for $i$-th star \\
$\bm{P}$ & covariance matrix \\
$\bm{Q}$ & process noise spectral density matrix \\
$\bm{R}$ & measurement covariance matrix \\
$\bm{R}_{bi}$ & DCM from body to inertial frame \\
$\bm{R}_{cb}$ & DCM from camera to body frame \\
$\alpha$ & right ascension [deg] \\
$\beta$ & magnitude of velocity ratio $v/c$ \\
$\bm{\beta}$ & velocity ratio vector $\bm{v}/c$ \\
$\gamma$ & Lorentz factor \\
$\delta$ & declination [deg] \\
$\delta\theta$ & parallax shift angle [arcsec] \\
$\delta\theta'$ & aberrated parallax shift angle [arcsec] \\
$\delta\bm{\theta}'$ & total angular shift vector [rad] \\
$\delta\bm{\theta}_a$ & aberration contribution [rad] \\
$\delta\bm{\theta}_p$ & parallax contribution [rad] \\
$\bm{\epsilon}$ & line-of-sight perturbation vector [rad] \\
$\eta_i$ & standard deviation of $i$-th star position [AU] \\
$\bm{\eta}$ & measurement noise vector \\
$\kappa$ & conditioning number\\
$\mu$ & Sun gravitational parameter [AU$^3$/day$^2$] \\
$\mu_\alpha$ & proper motion in right ascension [mas/yr] \\
$\mu_\delta$ & proper motion in declination [mas/yr] \\
$\bm{\omega}$ & acceleration process noise vector [AU/day$^2$] \\
$\rho_i$ & distance from spacecraft to $i$-th star [AU] \\
$\bm{\rho}_i$ & vector from spacecraft to $i$-th star [AU] \\
$\hat{\bm{\rho}}_i$ & unit vector from spacecraft to $i$-th star \\
$\hat{\bm{\rho}}_i'$ & aberrated unit vector \\
$\sigma$ & standard deviation of angular uncertainty [arcsec] \\[6pt]

\multicolumn{2}{@{}l}{\textbf{Subscripts}} \\
$i$ & star index \\
$k$ & time or measurement index \\
$0$ & reference or initial \\
$s$ & star \\
$c$ & camera frame \\
$g$ & gravity \\
$a$ & aberration \\
$p$ & parallax \\

\end{longtable*}}

\clearpage

\section{Introduction}

The outer solar system represents one of the main frontiers of space exploration. At the time of writing, the Voyager 1 and Voyager 2 missions, launched in 1977, have reached 168 AU and 140 AU from the Sun, respectively, after more than 48 years from launch~\cite{kohlhase1977voyager}. They are currently traveling at heliocentric velocities of approximately 3.6 AU/yr and 3.2 AU/yr, respectively. Both the Voyager probes entered the near interstellar space, surpassing the boundary of the heliosphere located at approximately 120 AU from the Sun along their trajectories~\cite{suess1990heliopause}. While the Sun’s gravity still holds to a distance in the order of $10^5$AU, as per the Hill sphere calculation, the region beyond the heliopause is already marked as interstellar owing to the different environment with respect to the interplanetary case. In this region, indeed, the solar wind gradually leaves room for the presence of the interstellar medium~\cite{cox2005three}. Past this point, the gravitational influence of the Sun remains significant but increasingly weak, giving way to perturbations from other stars and the Galactic tidal field. This region is sparsely populated by icy bodies as trans-Neptunian objects up to the inner Oort Cloud objects~\cite{weissman1990oort}, making this zone a transitional region between the Solar System and the deep interstellar space~\cite{rickman2008injection}. Recently, owing to advancements in rocket launchers and spacecraft propulsion systems, missions to Neptune, trans-Neptunian objects, and the outer solar system are under investigation~\cite{dachwald2005optimal, aime2021exploration}.

Navigation of spacecraft within the interplanetary part of the Solar System has traditionally relied on Earth-based radiometric tracking techniques, such as one-way and two-way ranging and Doppler measurements~\cite{thornton2003radiometric}. These methods exploit the transmission and reception of electromagnetic signals to infer spacecraft position and velocity relative to Earth, leveraging the signal time of flight for ranging and the Doppler shift content for radial velocity information. These measurements are processed through navigation filters at ground stations to reconstruct the spacecraft orbit~\cite{ely2022comparison}. {With the Delta-Differential One-Way Ranging (Delta-DOR) method, two ground stations are used to refine the spacecraft orbit and achieve sub-kilometre accuracies even at distances of Jupiter~\cite{james2009implementation}.} Alternative navigation methods for interplanetary navigation have also been proposed and experimented \cite{bhaskaran2000deep}, such as using sightings of planets and asteroids~\cite{broschart2019kinematic, krause2024lonestar, franzese2021deep, raymond2015interplanetary, andreis2022onboard, henry2023absolute, andreis2024autonomous}, the Sun \cite{franzese2025autonomous}, {stars \cite{christian2019starnav, melvin1996kalman},} or observation of pulsars~\cite{winternitz2018sextant, zheng2019orbit, anderson2015validation}. While accurate and effective for interplanetary missions, the extension of ground-based tracking to a large number of spacecraft in the outer solar system faces severe challenges. Signal propagation delays become significant: for each additional 100~AU of distance, for instance, a two-way communication for a single satellite requires an additional time of 1 day and 4 hours for a round trip. This is almost 3 days for a satellite at 250 AU. Also, signal strength decreases with the square of distance, rapidly demanding excessive onboard transmission power. Considering the current growth of satellites, ground-based tracking will become increasingly impractical due to long latency and unfeasible at large heliocentric distances due to power constraints. As more mission concepts targeting the outer Solar System~\cite{quarta2010electric} and even interstellar space are under investigation~\cite{parkin2018breakthrough}, the development of suitable autonomous navigation methods becomes essential to sustain the development of multiple and simultaneous missions. Such methods could serve as either backup solutions while waiting for ground-based position fixes or even as primary solutions along lasting cruising phases.

{This paper investigates the use of the parallactic shift of nearby stars for spacecraft navigation in the outer solar system at distances of up to 250 AU from the Sun. In this region, the parallactic shifts are found to be measurable with enough accuracy to allow a spacecraft to determine its own position vector to the Sun, and to estimate its velocity through sequential measurements. In this formulation, aberration and parallax shift effects are treated simultaneously for spacecraft position and velocity estimation. Note that, in the inner solar system, the aberration effect dominates the parallactic shift contribution, and it can be isolated. The spacecraft attitude is found to be recoverable as in the interplanetary case, since star pattern matching algorithms rely on distant stars with negligible parallax. This allows treating the nearby stars as 3D kinematic points useful for position triangulation, and distant stars as traditional 2D directional references for attitude determination.} Note that this approach is valid within the region of interest of this paper, as it will be shown. It is well known that in the deep interstellar space, many light-years away from the Sun, the traditional notion of star pattern descriptors, commonly used in star trackers for attitude determination, becomes largely inapplicable since all the stars need to be treated as 3D points~\cite{mckee2022navigation}. {The region of interest considered in this study spans heliocentric distances from 30~AU to 250~AU for several reasons. First, Voyager~1, the most distant operational spacecraft to date, is currently located at approximately 168~AU from the Sun. Extending the analysis to 250~AU therefore encompasses the foreseeable operational regime of Voyager~1 in the coming decades, as well as other outbound missions such as New Horizons, Voyager~2, and the Pioneer spacecraft. Second, at heliocentric distances beyond approximately 250~AU, an increasing number of stars exhibit significant parallactic signatures, which generally improves observability. Consequently, the most challenging regime for stellar parallax-based navigation lies closer to the Sun, where parallactic effects relative to a Sun-centered catalog are intrinsically smaller. Finally, recent advances in launch capabilities, particularly the availability of reusable launch systems, make it reasonable to anticipate an increasing number of missions targeting trans-Neptunian objects in the region beyond 30~AU, further motivating the selected distance range.} {Note also that X-ray pulsar navigation, with reported accuracies in the order of kilometers~\cite{sheikh2006spacecraft}, would be applicable in the same region. Still, it is worthwhile to investigate the use of nearby stars as either primary or additional navigation aids when the spacecraft is on long cruise phases and when high-precision navigation is less critical. Additionally, this method could also be employed in cases of the unavailability of X-ray sensors onboard a spacecraft.}

This paper is organized as follows. Section~\ref{sec:stars} discusses the stellar sources and the observation models used in this study. Section~\ref{sec:methodology} introduces and details the star-based navigation principles, which are then elaborated into a navigation filter in Section~\ref{sec:filter}. The method is evaluated under representative test cases in Section~\ref{sec:performances} to assess performance. Eventually, concluding remarks and findings of this investigation are summarized in Section~\ref{sec:conclusions}.

\section{Stellar Sources} \label{sec:stars}

Astrometric measurements are expressed within the International Celestial Reference System (ICRS), a quasi-inertial reference system with origin in the solar system barycentre (SSB) and defined by the positions of distant extragalactic radio sources with negligible proper motion~\cite{ma1998international}. Its realization, the International Celestial Reference Frame (ICRF), establishes the principal axes of the celestial reference system with an accuracy at the micro-arcsecond level~\cite{charlot2020third}. Stars are cataloged in one of the main ICRF frames, according to the epoch of observation and mission. Examples include the Hipparcos~\cite{van1997hipparcos} and Gaia~\cite{prusti2016gaia} missions, which produced star catalogs in the J1991.25 and J2016.0 frames, respectively. Hipparcos provided the first space-based astrometric catalog, containing $\sim$118,000 stars with typical positional accuracies of 1 milliarcsecond (mas), along with proper motion and parallax measurements~\cite{perryman1997hipparcos}. The Tycho-2 extension added over 2.5 million stars with lower accuracy~\cite{hog2000tycho}. The Gaia mission, through successive data releases, has provided astrometric solutions for $\sim$1.8 billion sources, with accuracies of 20–30 $\mu$as for stars brighter than mag 15 and including radial velocities for $\sim$33 million objects~\cite{vallenari2023gaia}. These catalogs are useful to retrieve the astrometric state of a star, which is generally described using a five- or six-parameter kinematic model. 

The five parameter model describes the 3D position and 2D transversal velocity of a star relative to the solar system barycenter, while the six parameter model also includes the radial velocity information. The angular position of a star is given by its right ascension ($\alpha$) and declination ($\delta$) referred to an ICRS system at a given epoch, while its distance is inversely proportional to the cataloged parallax ($p$). Catalogs also contain the star motion parameters in terms of radial velocity ($v_r$), proper motion in right ascension ($\mu_\alpha$), and proper motion in declination ($\mu_\delta$). These parameters constitute the full set of information required to propagate the motion of stars. The transformation from catalog astrometric parameters $(\alpha, \delta, p, \mu_\alpha, \mu_\delta, v_r)$ to Cartesian state vectors in terms of star position and velocity vectors is summarized in ~\ref{appendix:catalogue_transform}. Note that, traditionally, stellar motion is approximated with a constant velocity because accelerations and the related path deviations are small over centuries. This modeling is widely used in literature~\cite{mayor1974kinematics}. Note also that the star catalog positions are related to the barycentric geometric directions of incoming light rays at the SSB at a standard epoch, with aberration, parallax, and light deflection corrections removed. They do not apply the light-time correction for the star positions, since, otherwise, spacecraft would not be able to make direct comparisons with the catalog and would need to compute light-times for each star. In this way, spacecraft-specific parallax and aberration corrections can be applied through the solar system. Still, a minor delta light-time effect is present due to the spacecraft position with respect to the SSB. In general, any star line-of-sight (LoS) observation made by a spacecraft placed at a heliocentric position $\bm{r}$, with a velocity $\bm{v}$, and at a different epoch t, will contain shifts with respect to the catalog data. The parallax and delta light-time effects are due to $\bm{r}$, the aberration effect is due to $\bm{v}$, and the star kinematics is due to $t$. These effects are detailed in the following. 

\subsection{Time Variation}
Let us consider a fictitious stationary observer placed at the SSB. In this case, while there are no parallax, delta light-time, and aberration effects with respect to the cataloged information, the stars will move according to the proper motion parameters. This motion can be expressed in either spherical or Cartesian coordinates as detailed in ~\ref{appendix:catalogue_transform}. For the remainder of this paper, we will refer to the Cartesian formulation with the star kinematics modelled as
\begin{equation}
\bm{r}_i(t) = \bm{r}_i(t_0) + \bm{v}_i (t - t_0)
\end{equation}
where $\bm{r}_i$ denotes the heliocentric position vector of the $i$-th star, $\bm{v}_i$ its heliocentric velocity vector, $t_0$ denotes the catalog reference time and $(t - t_0)$ the elapsed time. The corresponding line-of-sight directions from the SSB to the star will evolve accordingly. Note that this position vector is along the light path travel to the star as seen from the SSB.

\subsection{Delta light-time effect}
Consider a fictitious stationary observer located at the spacecraft heliocentric position vector $\bm{r}(t_0)$ at the catalog reference epoch $t_0$. Let $\bm{r}_i(t_0)$ denote the catalog position vector of a star along the light path with respect to the Solar System Barycenter. In principle, there is a small difference in the starlight emission time if the star is observed at the SSB and at the spacecraft location. If the spacecraft is closer to the star, the light detected at $t_0$ was emitted at a slightly earlier epoch than for the SSB observer; conversely, if farther, at a slightly later epoch. Within 250~AU from the Sun, the maximum delta light-time is $250\,\mathrm{AU}/c \approx 1.44$~days, where $c$ is the speed of light. This in principle moves the reference position of the star, but this shift is negligible. Over this interval, indeed, given the typical stellar velocities and large distances (see~\ref{appendix:parallax}), the maximum angular effect on the apparent star position is for the fastest star in proximity to the Solar System, that is the Barnard’s star. For this star, the angular shift due to the delta light time is less than $0.041$ arcseconds. This is negligible with respect to the more marked geometric parallax induced by the spacecraft position, as it will be shown.

\subsection{Parallax Shift Effect} 
Consider a fictitious stationary observer located at the spacecraft heliocentric position vector $\bm{r}$ at a given reference epoch. Let $\bm{r}_i$ denote the position vector of a star along the light path with respect to the SSB at the same reference epoch. Considering the negligible delta light-time difference, we can define the star position vector as observed from the spacecraft, $\bm{\rho}_i$, from the simple triangle relation
\begin{equation} \label{eq:triangle}
\bm{\rho}_i = \bm{r}_i - \bm{r}
\end{equation}
The apparent parallax shift is the angle between the star direction as seen from the SSB and the one as observed by the spacecraft. This is given by
\begin{equation} \label{eq:parallax-angle}
\delta \theta = \arccos\!\left(\hat{\bm{\rho}}_i^\top \hat{\bm r}_i \right)
\end{equation}
where $\hat{\bm r}_i = \bm{r}_i/\|\bm{r}_i\|$ denotes the star unitary direction with respect to the SSB and $\hat{\bm{\rho}}_i = \bm{\rho}_i/\|\bm{\rho}_i\|$ denotes the star unitary direction as observed from the spacecraft. Within 250 AU of the Sun, we have that $\|\bm{r}\| \ll \|\bm{r}_i\|$, and this allows expanding the line-of-sight direction from the spacecraft to the star to first order as
\begin{equation} \label{eq:parallax-firstorder}
\hat{\bm\rho}_i 
= \frac{\bm{r}_i - \bm{r}}{\|\bm{r}_i - \bm{r}\|}
\approx \hat{\bm r}_i - \frac{(\bm I - \hat{\bm r}_i \hat{\bm r}_i^\top)\,\bm{r}}{r_i}
+ \mathcal{O}\!\left((r/r_i)^2\right)
\end{equation}
where $\|\bm{r}\| = r$ and $r_i = \|\bm{r}_i\|$. Eq.~\ref{eq:parallax-firstorder} has relied on the expansion of the denominator as 
\begin{equation}
\|\bm{r}_i - \bm{r}\| = \left\| r_i \left(\hat{\bm{r}}_i - \frac{\bm{r}}{r_i}\right)\right\| = r_i \, \sqrt{1 - 2 \, \hat{\bm{r}}_i^\top \hat{\bm{r}} \frac{r}{r_i} + \frac{r^2}{r_i^2}}
\approx r_i \left(1 - \hat{\bm{r}}_i^\top \hat{\bm{r}} \frac{{r}}{r_i}\right)
+ \mathcal{O}\!\left((r/r_i)^2\right)
\end{equation}

obtained considering the approximation $\sqrt{(1 + \epsilon)} \approx 1 + \epsilon/2 + \mathcal{O}(\epsilon^2)$. From this formulation, Eq. \ref{eq:parallax-firstorder} is derived considering that, for $\eta \ll 1$, we have ${(1 + \eta)}^{-1} \approx 1 - \eta$. From Equation~\eqref{eq:parallax-firstorder}, we can note that the parallax vanishes if $\bm{r}$ is aligned with $\hat{\bm r}_i$, as expected, while it is maximized when $\bm{r}$ is orthogonal to the star direction, also as expected. Note that the Eq.~\ref{eq:parallax-firstorder} describes a projection of $\bm{r}/r_i$ onto the plane orthogonal to $\bm{\hat{r_i}}$ through the projection matrix $\bm I - \hat{\bm r}_i \hat{\bm r}_i^\top$. The line-of-sight formulation in Eq. \ref{eq:parallax-firstorder} is useful as it is ready for implementation in a predictor/corrector navigation filter in the unknown $\bm{r}$. Note also that Eq.~\eqref{eq:parallax-firstorder} can be re-normalised to enforce unitary length.

It is useful to evaluate the parallax contributions in given cases. As an example, for transverse baselines of 75~AU, 150~AU, and 250~AU from the Sun, and for a star located 5 light-years away ($\approx316{,}205$~AU), the corresponding parallax shifts are approximately $49$ arcseconds, $98$ arcseconds, and $163$ arcseconds, respectively. Note that the induced approximation error due to the linearized formulation in Eq.~\eqref{eq:parallax-firstorder} remains below 0.065~arcseconds even for the largest baseline of 250~AU, which is negligible compared to the corresponding parallax magnitudes. {For reference, the effect of stellar parallax has already been measured onboard New Horizons~\cite{lauer2025demonstration}, which at the time of writing is at 62~AU from the Sun. The observations were carried out using the LORRI instrument~\cite{cheng2008long}, a visible-light telescope with a $1024\times1024$ pixel detector and a $0.29^\circ$ field of view. New Horizons has detected angular parallax shifts of Proxima Centauri and Wolf 359 through LORRI~\cite{lauer2025demonstration}. At a spacecraft distance of 47 AU, the New Horizons mission measured the parallactic displacement of Proxima Centauri as 32.4 arcseconds with respect to Earth-based measurements, and of Wolf 359 as 15.7 arcseconds. ~\ref{appendix:parallax} provides a list of parallax shifts for the closest stars to the solar system per different transversal baseline distances.}

\subsection{Aberration effect}

Consider now an observer located at the SSB but moving with velocity $\bm{v}$ relative to it. Since the observer is at the SSB, no parallactic shift due to position occurs; however, the observer’s velocity $\bm{v}$ and the finite speed of light $c$ introduce the stellar aberration effect. Note that astrometric parameters in the catalogs are given along the non-aberrated light paths to stars. Let $\bm{\hat \rho}_i$ denote the inertial line-of-sight unit vector along the photon path to a star, and let $\bm{\hat \rho}_i'$ denote the aberrated LoS direction to the same star. The exact aberrated direction $\bm{\hat \rho}_i'$ is obtained from the relativistic velocity-addition law as~\cite{shuster2003stellar}
\begin{equation} \label{eq:exact_aberration}
    \bm{\hat \rho}_i' =
    \frac{1}{\gamma\,\bigl(1 + \bm{\beta}^\top \bm{\hat \rho}_i\bigr)}
    \Bigl(\bm{\hat \rho}_i + \gamma\,\bm{\beta}
    + \frac{\gamma - 1}{\beta^2}\,(\bm{\beta}^\top \bm{\hat \rho}_i)\,\bm{\beta}\Bigr)
\end{equation}
where $\bm{\beta}$ is the observer velocity ratio and $\gamma$ the corresponding Lorentz factor defined as
\begin{equation}
    \bm{\beta} = \frac{\bm{v}}{c} \quad ; \quad
    \beta = \|\bm{\beta}\| \quad ; \quad
    \gamma = \frac{1}{\sqrt{1-\beta^2}}
\end{equation}
Equation~\eqref{eq:exact_aberration} is exact and preserves the unitary norm of the line-of-sight direction. For non-relativistic spacecraft velocities, $\|\bm{v}\|\ll c$, and typical deep-space spacecraft velocities are below the order $\beta\sim 10^{-4}$. Thus, within this paper, we can expand Eq.~\eqref{eq:exact_aberration} to first order in $\beta$. Note that, for spacecraft moving at non-relativistic velocities, the Lorentz factor can be expanded as
\begin{equation}
    \gamma = \frac{1}{\sqrt{1-\beta^2}} = 1 + \frac{1}{2}\beta^2 + \mathcal{O}(\beta^4)
\end{equation}
Hence, all terms involving $(\gamma - 1)$ are of order $\mathcal{O}(\beta^2)$, and can be neglected compared to first-order terms in $\beta$. Therefore, for the linearized treatment of stellar aberration, we can approximate $\gamma \simeq 1$ and retain only first-order terms in $\beta$. With this in hand, we can consider the expansion $(1+ \epsilon)^{-1} = (1 - \epsilon) + \mathcal{O}(\epsilon^2)$, where $\epsilon \ll 1$, leading to  
\begin{equation}
\frac{1}{\gamma \,(1+\bm{\beta}^\top\bm{\hat \rho_i})} = 1 - (\bm{\beta}^\top\bm{\hat \rho}_i) + \mathcal{O}(\beta^2)
\end{equation}
Therefore, neglecting terms of order $\mathcal{O}(\beta^2)$, Eq.~\eqref{eq:exact_aberration} becomes
\begin{equation}
\bm{\hat \rho}_i' = \bigl(1 - \bm{\beta}^\top\bm{\hat \rho}_i\bigr)\bigl(\bm{\hat \rho}_i + \bm{\beta}\bigr) = \bm{\hat \rho}_i + \bm{\beta} - (\bm{\beta}^\top\bm{\hat \rho_i})\,\bm{\hat \rho}_i
\end{equation}
which can be rearranged as
\begin{equation}\label{eq:lin_aberration}
\bm{\hat \rho}_i' = \bm{\hat \rho}_i + \frac{(\bm{I} - \bm{\hat \rho}_i\bm{\hat \rho}_i^\top)\,\bm{v}}{c}
\end{equation}
Equation~\eqref{eq:lin_aberration} shows that, for non-relativistic velocities, the aberration is the projection of $\bm{\beta}$ onto the plane orthogonal to $\bm{\hat \rho}_i$. For stars in the velocity direction, then $\bm{\hat \rho}_i'-\bm{\hat \rho_i}=\bm{0}$ and aberration vanishes, as expected. If $\bm{\hat \rho}_i$ is orthogonal to $\bm{\beta}$, the aberration reaches its maximum value. Note that the linearized $\bm{\hat \rho}_i'$ has unit norm up to $\mathcal{O}(\beta^2)$, and any norm error appears at second order ($10^{-8}$ difference for $\beta \sim 10^{-4}$). We can consider the actual heliocentric velocities of the Voyager and New Horizon missions to estimate typical aberration amounts for missions to the outer solar system. At the time of writing, the Voyager 1 and Voyager 2 missions are travelling at 16.9 km/s and 15.3 km/s, respectively, and New Horizon has a velocity of 13.6 km/s. The resulting beta factors are all lower than $5.64\cdot10^{-5}$, well within the $\beta \sim 10^{-4}$ assumption of this section. For the fastest of these missions, Voyager 1, the maximum amount of the aberration shift, for a star located orthogonally to the spacecraft velocity vector, is $\sim11.6$ arcsec. {Therefore, the aberration effect has to be considered for navigation, but it has generally a lower impact than the parallactic shift within the region of interest of this paper. Note that, in the inner solar system, aberration dominates the parallactic shift effect. In the attitude determination problem, aberration is typically accounted for as a post-analysis correction \cite{yoon2011new}.}

\subsection{Combined effect} \label{subsec:combined}

We can now consider the combined effect of parallax and aberration on the line-of-sight direction measurement to stars. The delta light-time to stars plays a negligible role within 250~AU from the Sun. Also, the gravitational light deflection by the Sun should be considered when an accuracy better than $100$ microarcseconds is desired, but such accuracies cannot be detected by common navigation sensors, and therefore, they also can be neglected.

Let us consider a spacecraft located at a heliocentric position vector $\bm{r}$ with a non-relativistic heliocentric velocity $\bm{v}$ at a coordinate time $t$. First, the star parameters have to be propagated from the catalog reference time $t_0$ to $t$ to ensure observations refer to the same epoch. Then, the LoS to a star as measured by the spacecraft will contain shifts with respect to the star catalog for parallax owing to $\bm{r}$ and aberration due to $\bm{v}$. Let us now denote the observed LoS direction to the star by the spacecraft as $\bm{\hat \rho}_i'$, which includes both parallax and aberration effects. The angular shift between this observed LoS direction and the cataloged star direction is denoted as $\delta\theta'$, and amounts to
\begin{equation} \label{eq:aberratedparallaxshift}
\delta \theta' = \textrm{acos} (\bm{\hat \rho}_i'^\top \bm{\hat r}_i)
\end{equation}
where $\bm{\hat r}_i$ is the catalog unit vector to the star. Provided a sufficient distance from the SSB and considering nearby stars, the overall shift is directly measurable from a spacecraft imaging sensor with a narrow field of view, as $\bm{\hat r}_i$ is obtained by the propagated catalog data, and $\bm{\hat{\rho}}_i'$ is measured by star centroiding and the camera model transformations. These transformations relate the detected star centroid in pixel coordinates to the LoS direction, which can be then translated into an inertial measurement through the spacecraft attitude (~\ref{app:pinhole}). Now, recall that the linearized aberration from Eq.~\eqref{eq:lin_aberration} reads
\begin{equation} \label{eq:aberrationeq}
\bm{\hat \rho}_i' = \bm{\hat \rho}_i + (\bm{I} - \bm{\hat \rho}_i \bm{\hat \rho}_i^\top)\,\frac{\bm{v}}{c}
\end{equation}
and the first-order parallax expansion from Eq.~\eqref{eq:parallax-firstorder} is
\begin{equation} \label{eq:parallax-firstorder-restated}
\bm{\hat\rho}_i = \bm{\hat r}_i - \frac{\bigl(\bm{I}-\bm{\hat r}_i\bm{\hat r}_i^{\!\top}\bigr)\,\bm r}{r_i}
\end{equation}
In Eq.~\eqref{eq:aberrationeq}, the projector $(\bm{I} - \bm{\hat \rho}_i \bm{\hat \rho}_i^\top)$ ensures that only the angular component of the velocity shift contributes perpendicular to the instantaneous LoS. Substituting the first-order parallax expansion, that is Eq.~\eqref{eq:parallax-firstorder-restated} into Eq.~\eqref{eq:aberrationeq}, generates higher-order terms of the form $(\bm{I} - \bm{\hat \rho}_i \bm{\hat \rho}_i^\top)\,\bm r^\top\bm{v}/ (r_i \,c)$, which are second-order in the small parameters $\beta = v/c$ and $r/r_i$. Since $\beta \lesssim 10^{-4}$ and $r/r_i \lesssim 10^{-7}$, these contributions are well below 0.1 arcseconds and therefore can be neglected. Therefore, to first order, we can replace $\bm{\hat \rho}_i$ by $\bm{\hat r}_i$ inside the projectors, yielding the compact vector expression for the observed line-of-sight direction as
 
\begin{equation} \label{eq:compact-vector}
\bm{\hat\rho}_i' = 
\bm{\hat r}_i 
+ \bigl(\bm{I}-\bm{\hat r}_i\bm{\hat r}_i^{\!\top}\bigr)\!\left(\frac{\bm{v}}{c} - \frac{\bm r}{r_i}\right) 
\end{equation}
Note that this equation is useful as it is ready for implementation into a navigation filter in both the spacecraft state variables $\bm{r}$ and $\bm{v}$, with all other parameters known. For a further analysis, we can identify the small angular shift vector as
\begin{equation} \label{eq:delta-vector}
\delta\bm\theta' = \bm{\hat\rho}_i' - \bm{\hat r}_i
= \bigl(\bm{I}-\bm{\hat r}_i\bm{\hat r}_i^{\!\top}\bigr)\!\left(\frac{\bm{v}}{c} - \frac{\bm r}{r_i}\right)
\end{equation}
In this, we can isolate the two angular shift contributions as $\delta\bm\theta_{\rm p}$ due to parallax and $\delta\bm\theta_{\rm a}$ due to aberration, respectively, as 
\begin{equation}
\delta\bm\theta_{\rm p} = -\,\bigl(\bm{I}-\bm{\hat r}_i\bm{\hat r}_i^{\!\top}\bigr)\,\frac{\bm r}{r_i} \quad ; \quad
\delta\bm\theta_{\rm a} = \bigl(\bm{I}-\bm{\hat r}_i\bm{\hat r}_i^{\!\top}\bigr)\,\frac{\bm{v}}{c}
\end{equation}
Note that the norm $\delta\theta' = \|\delta\bm\theta'\|$ can be linked to Eq.~\ref{eq:aberratedparallaxshift} through the small-angle expansion

\begin{equation} \label{eq:delta-scalar}
\textrm{cos} \, \delta\theta' \approx 1 - \frac{1}{2} \, \delta\theta'^2
\end{equation}

\begin{figure}[htbp]
  \centering
  \begin{subfigure}[b]{0.33\textwidth}
    \centering
    \includegraphics[width=\linewidth]{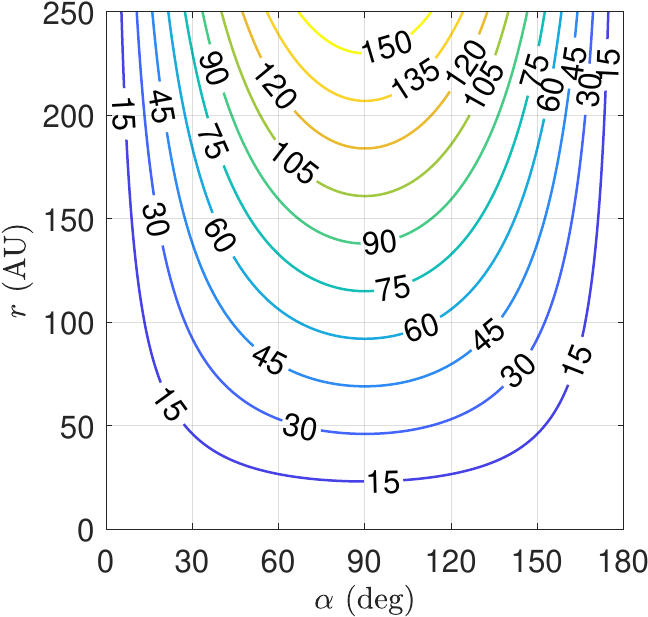}
    \caption{Parallax shift for stars located at 5 light-years as a function of the angle between the spacecraft position and star direction.}
    \label{fig:parallax5ly}
  \end{subfigure}
  \hfill
  \begin{subfigure}[b]{0.33\textwidth}
    \centering
    \includegraphics[width=\linewidth]{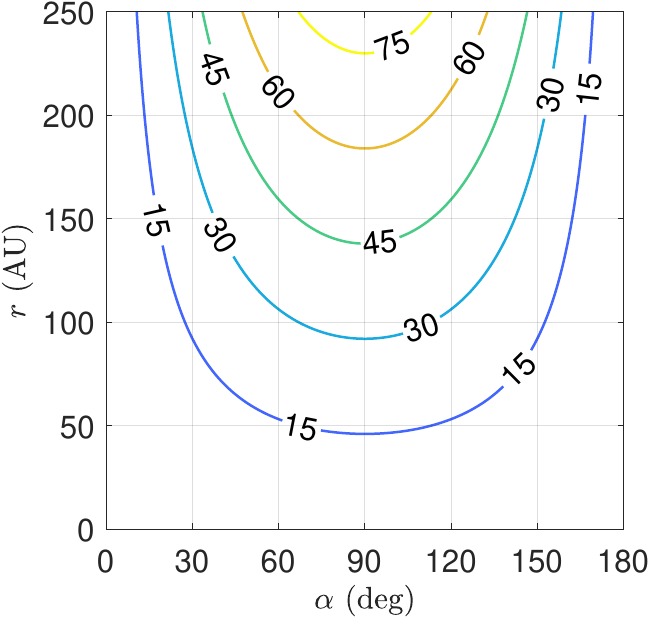}
    \caption{Parallax shift for stars located at 10 light-years as a function of the angle between the spacecraft position and the star direction.}
    \label{fig:parallax10ly}
  \end{subfigure}
  \hfill
  \begin{subfigure}[b]{0.32\textwidth}
    \centering
    \includegraphics[width=\linewidth]{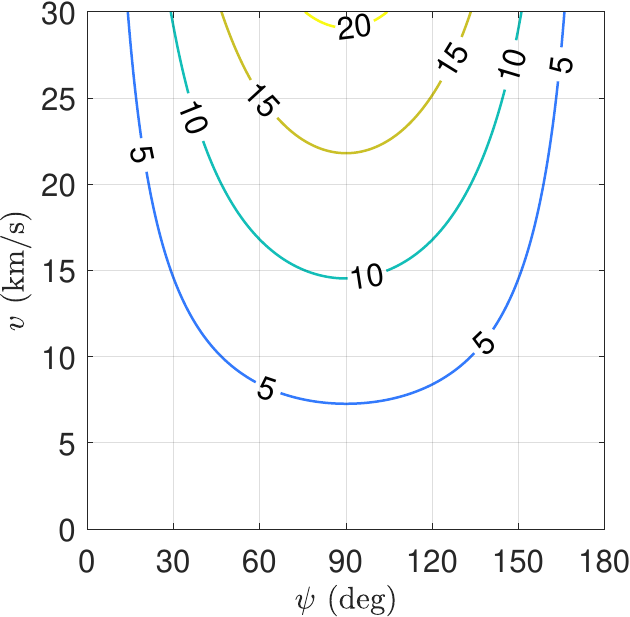}
    \caption{Aberration of stars as a function of the spacecraft velocity and angle between the velocity vector and star direction.}
    \label{fig:aberration}
  \end{subfigure}
  \caption{Parallactic shift of nearby stars located at (a) 5 ly and (b) 10 ly, considering spacecraft positions within 250 AU from the Sun, and (c) aberration effect for velocities up to 30 km/s ($\beta \approx 10^{-4}$). The angle between the heliocentric spacecraft position vector and the star direction with respect to the SSB is denoted $\alpha$, while the angle between the velocity vector and the star direction is denoted $\psi$. Parallax and aberration values given in arcseconds.}
  \label{fig:parallax-aberration}
\end{figure}

Without an a priori coarse knowledge of the spacecraft position and velocity vectors, however, the parallax and aberration contributions to the line-of-sight angular shift cannot be distinguished. Still, even with large uncertainties while estimating $\bm{r}$ and $\bm{v}$, as in a filter loop, they can be well separated. It is therefore worthwhile to evaluate their relative magnitudes. Figure~\ref{fig:parallax-aberration} shows the parallax and aberration contributions to the angular shift of the line-of-sight directions. The parallax contribution is plotted as a function of the spacecraft heliocentric distance and angle between the spacecraft position vector and the star position vector in the SSB frame, for stars located at 5~ly and 10~ly in Figure \ref{fig:parallax5ly} and Figure \ref{fig:parallax10ly}, respectively. The aberration has been computed for $\beta$ values up to $10^{-4}$, considering the angle between the spacecraft velocity vector and the line-of-sight to the star. As it can be seen, the parallax contribution dominates the angular shift effect for non-relativistic spacecraft within 250~AU from the Sun. Still, aberration plays a non marginal role, and it must be considered. Other effects, such as delta light-time and gravitational light deflection by the Sun are orders of magnitude lower, and can be neglected. Note that values in Fig.~\ref{fig:parallax-aberration} have been computed using the linearized LoS direction formulas. For validation, these values have also been computed using the exact parallax and aberration in Eq.~\eqref{eq:parallax-angle} and Eq.~\eqref{eq:exact_aberration}, respectively, finding a maximum relative error between the linearized and exact formulations of $0.05\%$ regarding parallax and $0.01\%$ regarding aberration. The absolute error values of these contributions in the region of interest of this paper are lower than 0.09 arcseconds. This validates using the linearized LoS directions, noting that these formulations are also useful for implementation into navigation filters. 

\section{Position Estimation} 
\label{sec:methodology}

Section \ref{sec:stars} has shown that, within 250~AU from the Sun and for spacecraft velocities with $\beta \le 10^{-4}$, the parallactic shift effect becomes evident for nearby stars, and the aberration plays a non-negligible role. Within this regime, we can note that the spacecraft attitude can be recovered using conventional star trackers and state-of-the-art methods, as in the interplanetary case, by exploiting distant stars and established star pattern descriptors~\cite{wertz2012spacecraft}. This is because typical star catalogs loaded in star trackers exclude nearby stars to avoid parallax effects. As an example, many star trackers do not list stars closer than 100 light-years, as evidenced by Shuster~\cite{shuster2003stellar}. If such stars are actually listed, we can consider excluding them from the star tracker catalog, without any effect on attitude observability, given the enormous amount of stars contained in star tracker catalogs. {For instance, consider a common star tracker with 12° FoV and 1024x1024 pixel detector, for which each pixel spans 42.18 arcseconds. Star trackers typically operate out of focus to spread the light of stars in several pixels and perform centroiding, thus achieving sub-pixel accuracy for centroid determination~\cite{rufino2003enhancement}. Let us consider a high-performance centroid determination accuracy of 0.25 pixels, which corresponds to 10.29 arcseconds for the considered star tracker. A star located 100 light years away produces a maximum parallax shift of 8.1 arcseconds at a transversal baseline of 250 AU, which is smaller than the centroiding accuracy of the considered star tracker.} Also, note that star pattern descriptors rely on multiple inter-star angle geometries, which span several degrees across the sky. A small offset in the order of arcseconds for the stars line-of-sight determination does not impact attitude observability, and still allows accurate attitude determination. State-of-the-art star trackers for traditional satellites typically achieve attitude determination accuracies between 1 and 5~arcseconds in a 3$\sigma$ standard deviation confidence along each axis~\cite{markley2014fundamentals}, relying on multiple optical heads pointing in different directions to detect different sets of stars. Note that even more accurate attitude determination performances can be reached, as the sub-arcsecond attitude reconstruction for the Gaia mission \cite{risquez2013attitude}.

\subsection{Angular shift detection} \label{sec:angular_shift}

The angular shift of stars close to the solar system, which is in the order of tens of arcseconds within a baseline of 250 AU, is well contained within the field of view of typical navigation cameras, which is in the order of degrees. Therefore, still with its position and velocity vectors unknown, but with attitude recovered, a spacecraft can slew and point towards the cataloged direction of these stars, aiming for an angular shift detection due to parallax. {As noted, the observability of the stellar parallax depends upon the sensor used and its ability to detect small angular shifts, apart from having a baseline transversal distance from the SSB. While for attitude determination we can consider star trackers with a relatively large FoV (e.g., $\ge$ 10 degrees) to maximize the number of observed stars and detect the inter-star angles, for position determination we need to consider a narrow FoV imaging camera to detect the small angular shifts of selected nearby stars. To this aim, we need to define a conservative threshold above which the angular shift is due to the observer position, and not due to sensor noise. Therefore, we will consider a robust center-to-center distance of at least two pixels between the measured star direction and the corresponding cataloged one, to not confuse a parallax effect with other effects. As an example, we can consider a narrow-FoV navigation camera with 2 deg FoV and 1024x1024 pixels detector, for which each pixel spans 7.03 arcseconds. The resulting two-pixels threshold is set to 14.06 arcseconds. This also marks the difference between nearby and distant stars. Nearby stars are those ones for which parallax can be measured according to the angular resolution of an imaging instrument, while distant stars are those ones that do not exhibit parallactic shift according to the imaging instrument angular resolution. Table~\ref{tab:parallax_2deg} describes the angular shifts for selected star distances and the considered baselines, along with their measurability according to the defined camera specifications and threshold. By scanning Table~\ref{tab:parallax_2deg}, we can note that the parallax of stars close to the solar system is already beyond the two-pixels threshold, and therefore, it can be considered measurable. This example is just one way of selecting a navigation camera for parallax detection. Engineers and navigation analysts are free to select the most adequate star tracker/nav-cam pair for the intended attitude determination and position determination problems within the region of this paper. Note that imaging instruments with a narrow FOV would be capable of detecting such angular shifts even at closer distances to the Sun, as is the case of the LORRI instrument ($0.29^\circ$ FOV) onboard New Horizons~\cite{weaver2020flight}.} 

\begin{table}[ht]
\centering
\caption{Angular shifts of stars and two-pixels detectability threshold for a camera with $2^\circ$ FoV and $1024\times1024$ pixels. The columns $\Delta \theta_{30}$, $\Delta \theta_{75}$, $\Delta \theta_{150}$, and $\Delta \theta_{250}$ denote the apparent angular shift of the star for a spacecraft baseline transversal displacement of 30 AU, 75 AU, 150 AU, and 250 AU, respectively. If the star angular shift is detectable and exceeding a defined threshold due to parallax, such a star can be used for spacecraft position estimation.}
\begin{tabular}{cccccccccc} \hline \hline
Distance & Distance & $\Delta \theta_{30}$ & Detectable & $\Delta \theta_{75}$ & Detectable & $\Delta \theta_{150}$ & Detectable & $\Delta \theta_{250}$ & Detectable \\
(ly) & (AU) & (arcsec) & & (arcsec) & & (arcsec) & & (arcsec) & \\ \hline
5    & 316,205   & 19.6   & Yes & 48.9    & Yes & 97.8    & Yes & 163.1   & Yes \\
10   & 632,410   & 9.8    & No  & 24.5    & Yes & 48.9    & Yes & 81.5    & Yes \\
20   & 1,264,820 & 4.9    & No  & 12.2    & No  & 24.5    & Yes & 40.8    & Yes \\
50   & 3,162,050 & 2.0    & No  & 4.9     & No  & 9.8     & No  & 16.3    & Yes \\
100  & 6,324,100 & 1.0    & No  & 2.4     & No  & 4.9     & No  & 8.2     & No  \\ \hline \hline
\end{tabular}
\label{tab:parallax_2deg}
\vspace{-0.25cm}
\end{table}

For reference, ~\ref{appendix:parallax} provides the catalog information and the computed angular shifts $\Delta \theta$ for the stars within 15 light years of the solar system, sorted for increasing distance to the SSB. By scanning the table, we can see that angular shifts of nearby stars are measurable given the assumed instrument and distances. Also note that the apparent magnitude of these nearby stars is well within typical limit magnitude for detection of common spacecraft imaging instruments~\cite{geiger2021radiometric}.
\subsection{Least Squares Estimation} \label{sec:positionEstimation}

This section describes a position estimation method with simultaneous measurements to N stars in a least squares sense. This method is useful to gather insights into the geometry of the problem, but it is impractical to gather simultaneous measurements to more than one star with parallax effect. This is because a spacecraft would need many narrow FoV cameras onboard to capture more than one star with apparent parallax. Another method, detailed in Section \ref{sec:filter}, considers just one star with apparent parallax tracked per time within a navigation filter. This is more prone to implementation, as it will be shown, as it allows the spacecraft to point and track multiple stars in a sequential way along a deep-space trajectory to estimate its orbit. 

Let us proceed with the least squares method first. Let us consider a spacecraft at unknown heliocentric position $\bm{r}$ and velocity $\bm{v}$ which seeks to navigate acquiring nearby stars. We will assume that the spacecraft can acquire its attitude through distant stars and point towards the cataloged directions of nearby stars for angular shift detection as described in the previous sections. At this stage, this shift cannot be divided into the parallax and aberration contributions, as both $\bm{r}$ and $\bm{v}$ are still unknown. However, after coarse estimations, they can be separated. Let us assume that the spacecraft is capable of detecting multiple line-of-sight directions to nearby stars with an apparent angular shift with respect to the cataloged parameters. Also, let us neglect the effects of delta light-time and gravitational light deflection. Denoting by $\hat{\bm{\rho}}_i$ the non-aberrated inertial line-of-sight unit vector pointing from the spacecraft to the $i$-th nearby star, we can write the spacecraft position as
\begin{equation} \label{eq:positionvector2}
\bm{r} = \bm{r}_i - \rho_i\, \bm{\hat\rho}_i \qquad i=1,\dots,N
\end{equation}
Eq.~\eqref{eq:positionvector2} is written as a function of the \(i\)-th star position, and it considers the parallax effect in $\bm{\hat\rho}_i$ due to $\bm{r}$. The unknowns are the spacecraft position vector $\bm{r}$ and the distance to each star $\rho_i$. We can project Eq.~\eqref{eq:positionvector2} onto the plane orthogonal to $\hat{\bm{\rho}}_i$ through its projection matrix $\bm{L}_i$, considering that
\begin{equation} \label{eq:positionvector4}
\qquad \bm{L}_i = \bm{I} - \hat{\bm\rho}_i\hat{\bm\rho}_i^\top \qquad ; \qquad \bm{L}_i \, \hat{\bm\rho}_i = \bm{0}\qquad i=1,\dots,N 
\end{equation}
Pre-multiplying both sides of Eq.~\eqref{eq:positionvector2} by $\bm L_i$ leads to
\begin{equation} \label{eq:positionvector3}
\bm{L}_i \, \bm{r} = \bm{L}_i \, \bm{r}_i \qquad i=1,\dots,N
\end{equation}
We can now stack the N equations from Eq.~\eqref{eq:positionvector3} to obtain the system
\begin{equation} \label{eq:stacked_projection}
\underbrace{\begin{bmatrix}
\bm{L}_1 \\[-6pt]
\vdots \\[-6pt]
\bm{L}_N
\end{bmatrix}}_{\bm{H}}
\bm{r} =
\underbrace{\begin{bmatrix}
\bm{L}_1 \bm{r}_1 \\[-6pt]
\vdots \\[-6pt]
\bm{L}_N \bm{r}_N
\end{bmatrix}}_{\bm{d}}
\end{equation}
where $\bm{H}$ has dimension 3N$\times$3 and $\bm{d}$ has dimension 3N$\times$1. Eq.~\eqref{eq:stacked_projection} is an overdetermined linear system in the unknown spacecraft position vector $\bm{r}$, which can be solved in a least-squares sense as
\begin{equation} \label{eq:least_squares_r}
\bm{r} = \left( \bm{H}^\top \bm{H} \right)^{-1} \bm{H}^\top \bm{d}
\end{equation}
Once $\bm{r}$ has been estimated from Eq.~\eqref{eq:least_squares_r}, if desired, the distance to each star can be obtained as
\begin{equation} \label{eq:rho_estimation}
\rho_i = \| \bm{r}_i - \bm{r} \| , \qquad i = 1, \dots, N
\end{equation}

Note that the least squares formulation in Eq.~\eqref{eq:least_squares_r} has not modeled the aberration effect in the line-of-sight directions to stars. However, aberration is present when a spacecraft measures the LoS directions. Therefore, this effect can be accounted for as a post-estimate correction. With sequential estimates $\bm{r}(t)$, a coarse velocity $\bm{v}(t)$ can be estimated. This can be used to refine the directions to the stars inverting Eq.~\eqref{eq:lin_aberration} and re-running the least squares to iterate the spacecraft position. When the spacecraft measures an aberrated direction $\bm{\hat \rho}_i'$ and has an estimate of $\bm{v}$, the non-aberrated direction $\bm{\hat \rho}_i$ can be reconstructed as
\begin{equation}\label{eq:lin_aberration_inverse}
\bm{\hat \rho}_i = \bm{\hat \rho}_i' - \frac{(\bm{I} - \bm{\hat \rho}_i'\bm{\hat \rho}_i'^\top)\,\bm{v}}{c}
\end{equation}
and the position estimation can be re-runned accordingly. Now, considering that each matrix $\bm{L}_i$ is symmetric and idempotent ($\bm{L}_i = \bm{L}_i^\top = \bm{L}_i^2$), we can note that the matrix $\bm{H}^\top\bm{H}$ can be written as
\begin{equation}
\bm{H}^\top \bm{H} = \sum_{i=1}^{N} \bm{L}_i = N\bm I_3 - \underbrace{\sum_{i=1}^N \hat{\bm\rho}_i\hat{\bm\rho}_i^\top}_{\bm S}
\end{equation}
A solution to Eq.~\eqref{eq:least_squares_r} exists provided $\sum_{i=1}^{N}\bm{L}_i$ is non-singular, which requires the line-of-sight directions $\hat{\bm{\rho}}_i$ not to be all coplanar or nearly parallel, and a minimum of two non-collinear directions \(\hat{\bm\rho}_i\) is required, as it is known in other estimation problems relying on line-of-sight directions \cite{franzese2022deep}. {The problem is well-conditioned when the directions are spread isotropically. For example, with \(N=2\) and an inter-angle \(\gamma\) such that \(\hat{\bm\rho}_1^\top\hat{\bm\rho}_2=\cos\gamma\), the eigenvalues of \(\bm H^\top\bm H\) are
\[
\mu_1 = 1 - \cos\gamma \qquad ; \qquad
\mu_2 = 1 + \cos\gamma \qquad ; \qquad
\mu_3 = 2
\]
and the conditioning number of $\bm{H}^\top\bm{H}$ is
\begin{equation}
\kappa(\gamma)=\frac{\max\{\mu_j\}}{\min\{\mu_j\}}=\frac{2}{1-|\cos\gamma|}
\end{equation}
which tends to infinity as \(\gamma\to 0^\circ\) and equals \(2\) at \(\gamma=90^\circ\). In the ideal isotropic limit for many directions one has \(\bm S\approx \tfrac{N}{3}\bm I_3\) and \(\bm H^\top\bm H\approx \tfrac{2N}{3}\bm I_3\), giving a well-conditioned matrix. Therefore, the problem is well conditioned when stars are spread apart, and this angular spread value is function of the number of stars considered. Note that this result, which comes from the least squares formulation, is also applicable to other estimators when observing a multitude of stars.}

\subsection{Position covariance}
We can consider uncertainties in both the star ephemerides and measurement noise of the line-of-sight directions and evaluate their impact on the least-squares position estimation. While uncertainties in star ephemerides can be retrieved by catalogs, those into line-of-sight direction depend mainly on the spacecraft attitude knowledge and image processing (e.g., centroiding) performance. Regarding star ephemerides, we can consider the uncertainty $\bm{\Delta r}_i$ in their position as
\begin{equation} 
\bm{r}_i^{\epsilon} = \bm{r}_i + \Delta \bm{r}_i
\end{equation}
where we can use the isotropic ephemeris uncertainty model
\begin{equation}
E[\Delta \bm{r}_i] = \bm{0} \qquad ; \qquad E[\Delta \bm{r}_i \Delta \bm{r}_i^\top] = \eta_i^2 \bm{I}
\end{equation}
with $\eta_i$ being the standard deviation in position retrieved by star catalogs. For the noisy line-of-sight directions $\hat{\bm\rho}^\epsilon_i$, given the small perturbation assumption, we can consider the QUEST measurement model (~\ref{appendix:quest}) where
\begin{equation}
\hat{\bm\rho}^\epsilon_i \approx \hat{\bm\rho}_i + \bm{v}_i   
\end{equation}
where $\bm{v}_i$ is a small perturbation orthogonal to $\hat{\bm\rho}_i$, for which
\begin{equation}
E[\bm{v}_i] = \bm{0} \qquad ; \qquad E[\bm{v}_i \bm{v}_i^\top] = \sigma_i^2 \bm{L_i}
\end{equation}
with $\sigma_i$ the standard deviation in angular uncertainty of the inertial line-of-sight direction. Considering the perturbed direction, we can write the perturbed projected equation as
\begin{equation} \label{eq:perturbationL}
(\bm{L}_i + \delta\bm L_i) \, \bm r = (\bm{L}_i + \delta \bm L_i) \, (\bm r_i + \Delta \bm{r_i})
\end{equation}
where we can derive the first-order perturbation \(\delta\bm L_i\) expanding \(\hat{\bm\rho}_i^\epsilon\) with respect to \(\hat{\bm\rho}_i\), leading to
\begin{equation}
\delta\bm L_i \approx -\big( \bm v_i\,\hat{\bm\rho}_i^\top + \hat{\bm\rho}_i \, \bm v_i^\top \big)
\end{equation}
where second-order terms have been neglected. In this way, Eq.~\ref{eq:perturbationL} can be developed as 
\begin{equation} \label{eq:noisy_ls}
\bm L_i \bm r = \bm L_i \bm r_i + \bm L_i\,\Delta\bm r_i
\;+\delta\bm L_i(\bm r_i - \bm r)
\end{equation}
Using \(\bm r_i - \bm r = \rho_i \hat{\bm\rho}_i\) and the first-order expression for \(\delta\bm L_i\), the directional term reduces to
\begin{equation}
\delta\bm L_i(\bm r_i - \bm r) = -\rho_i\,\bm v_i
\end{equation}
since \(\hat{\bm\rho}_i^\top \bm v_i=0\). Therefore, the noise on each block \(i\) of Eq.~\eqref{eq:noisy_ls} reads
\begin{equation}\label{eq:nu_block}
\bm w_i \approx \bm L_i\,\Delta\bm r_i \;-\; \rho_i\,\bm v_i
\end{equation}
which has contributions due to ephemeris uncertainty and angular LoS uncertainty. The block covariance of \(\bm w_i\) is
\begin{equation} \label{eq:nu_cov_block}
\mathbb{E}[\bm w_i \bm w_i^\top] \approx \mathbb{E}[\bm L_i\Delta\bm r_i\Delta\bm r_i^\top \bm L_i^\top] + \rho_i^2\,\mathbb{E}[\bm v_i\bm v_i^\top] = \eta_i^2 \bm L_i + \rho_i^2\sigma_i^2 \bm L_i = \big(\eta_i^2 + \rho_i^2\sigma_i^2\big)\,\bm L_i = \bm{R}_i
\end{equation}
Note that cross-terms vanish because independent each other. Now we can stack Equations~\eqref{eq:noisy_ls} leading to 
\begin{equation} \label{eq:stacked_projection_noisy}
\underbrace{\begin{bmatrix}
\bm{L}_1 \\[-6pt]
\vdots \\[-6pt]
\bm{L}_N
\end{bmatrix}}_{\bm{H}}
\bm{r} =
\underbrace{\begin{bmatrix}
\bm{L}_1 \bm{r}_1 \\[-6pt]
\vdots \\[-6pt]
\bm{L}_N \bm{r}_N
\end{bmatrix}}_{\bm{d}} +
\underbrace{\begin{bmatrix}
\bm{w}_1 \\[-6pt]
\vdots \\[-6pt]
\bm{w}_N
\end{bmatrix}}_{\bm{w}}
\end{equation}
where $\bm{w}$ is the stacked noise. The full stacked noise covariance, $\bm R$, is therefore the block-diagonal matrix
\begin{equation}
\bm R = \mathbb{E}[\bm w \bm w^\top] \approx
\mathrm{blockdiag}\left(\bm R_1, \dots, \bm R_N\,\right)
\end{equation}
Therefore, the position covariance $\bm{P}_r$ in the least squares estimation can be obtained as 
\begin{equation}\label{eq:cov_unweighted}
\bm{P}_r = (\bm H^\top\bm H)^{-1}\,\bm H^\top\bm R\,\bm H\,(\bm H^\top\bm H)^{-1}
\end{equation}
Substituting \(\bm H\) as the stacked \(\bm L_i\) and using \eqref{eq:nu_cov_block} yields the compact form
\begin{equation}\label{eq:cov_unweighted_sum}
\bm{P}_r = (\bm H^\top\bm H)^{-1}\left(\sum_{i=1}^N (\eta_i^2 + \rho_i^2\sigma_i^2)\,\bm L_i \right)
(\bm H^\top\bm H)^{-1}
\end{equation}
Note that the two contributions to the block variance are \(\eta_i^2\), which is the ephemeris position uncertainty projected into the tangent plane, and \(\rho_i^2\sigma_i^2\), that is the angular measurement error mapped into position through \(\rho_i\). In general, \(\rho_i\sigma_i \gg \eta_i\) for stars, and therefore, improving centroiding/attitude knowledge reduces position error. Note also that this derivation is first-order in \(\bm v_i\) and \(\Delta\bm r_i\) and neglects the influence of \(\delta\bm L_i\) on the geometry matrix beyond the additive measurement noise. For moderate angular errors in arcseconds the linear model is adequate. For larger errors a non-linear model is recommended.

\section{Navigation Filter} \label{sec:filter}
Section \ref{sec:methodology} has shown that the parallax of nearby stars can be exploited for position navigation, with an investigation of the impact of the observation geometry. This section details an extended Kalman filter (EKF) formulation \cite{kalman1960} that sequentially processes one star line-of-sight measurement at a time to estimate the spacecraft trajectory, since it is impractical to acquire more than one star with evident parallax per navigation window. The Kalman filter adopts prediction and correction steps. During the prediction stage the state and covariance are propagated to the next measurement time \(t_k\). The propagated state and covariance are denoted \(\bm x_k^-\) and \(\bm P_k^-\), respectively. After the correction step the posterior quantities are \(\bm x_k^+\) and \(\bm P_k^+\), respectively. This approach is discussed in the following.

\subsection{Propagation}
We can define the orbital state vector as
\begin{equation}
\bm x = \begin{bmatrix}\bm r \\[-6pt] \bm v\end{bmatrix}
\end{equation}
where \(\bm r\) and \(\bm v\) are the heliocentric inertial position and velocity vectors of the spacecraft expressed in the Sun-centered J2000 frame. The acceleration acting on the spacecraft is modelled as the sum of solar gravity and solar radiation pressure (SRP) as
\begin{equation}
\bm a = \bm a_g + \bm a_s
\end{equation}
with gravity acceleration $\bm a_g = -\mu r^{-3}\bm r$ and, assuming a simple cannonball SRP model,
\begin{equation}
\bm a_s = c_r \frac{S_0 d^2}{c}\frac{A}{m}\,\frac{\bm r}{r^3}
\end{equation}
In the previous, \(\mu\) is the Sun gravitational parameter, \(S_0\) the solar constant at 1~AU, \(d\) the mean Earth--Sun distance (1~AU), \(c\) the speed of light, \(A/m\) the area-to-mass ratio, and \(c_r\) the spacecraft reflectivity coefficient. The continuous-time dynamics including process noise can be written as
\begin{equation}\label{eq:system_corrected}
\dot{\bm x} = \bm f(\bm x,t) + \tilde{\bm w}
\quad\rightarrow\quad
\begin{bmatrix}\dot{\bm r}\\[-6pt]\dot{\bm v}\end{bmatrix}
=
\begin{bmatrix}\bm v\\[-6pt]\bm a\end{bmatrix}
+
\begin{bmatrix}\bm 0\\[-6pt]\bm\omega\end{bmatrix}
\end{equation}
where \(\tilde{\bm w}=[\bm 0^\top,\bm\omega^\top]^\top\) models continuous-time acceleration noise. The state error covariance \(\bm P=\mathbb E[\Delta\bm x\,\Delta\bm x^\top]\) propagates according to the continuous-time Riccati equation
\begin{equation}\label{eq:riccati}
\dot{\bm P} = \bm F\,\bm P + \bm P\,\bm F^\top + \bm Q
\end{equation}
where \(\bm F=\partial\bm f/\partial\bm x\) is the Jacobian of the dynamics
\begin{equation}
\bm F =
\begin{bmatrix}
\bm 0_3 & \bm I_3\\[-6pt]
\bm F_r & \bm 0_3
\end{bmatrix}
\end{equation}
and \(\bm F_r=\partial \bm a/\partial \bm r\) can be written in compact form as
\begin{equation}\label{eq:Fr}
\bm F_r
= \left(-\mu + c_r\frac{S_0 d^2}{c}\frac{A}{m}\right)\frac{\,\bm I_3 r^2 - 3\bm r\bm r^\top\,}{r^5}
\end{equation}

Note that \(\bm Q\) is the continuous-time process-noise spectral density. In block form consistent with Eq.~\eqref{eq:system_corrected}, one can write
\begin{equation}
\bm Q = \begin{bmatrix}\bm 0_3 & \bm 0\\[-6pt]\bm 0 & \bm Q_a\end{bmatrix}
\end{equation}

where \(\bm Q_a\) is a \(3\times3\) positive semi-definite matrix characterizing acceleration noise. Given an initial state $\bm{x}_0$ or previous updated state $\bm{x}_{k-1}^+$, the propagated state $\bm{x}_k^-$ and covariance $\bm{P}_k^-$ are obtained by integration Eq.~\eqref{eq:system_corrected} and Eq.~\eqref{eq:riccati}, respectively.

\subsection{Update} \label{subsec:Update}
The filter uses LoS measurements to catalogued stars which already account for parallax and aberration effects. From Section \ref{subsec:combined}, this LoS can be written as
\begin{equation}\label{eq:los_approx}
\hat{\bm\rho}_i' \approx \hat{\bm r}_i
+ \big(\bm I_3-\hat{\bm r}_i\hat{\bm r}_i^\top\big)\!\left(\frac{\bm v}{c}-\frac{\bm r}{r_i}\right)
\end{equation}
where \(\hat{\bm r}_i=\bm r_i/\|\bm r_i\|\) is the unit vector from the Sun to the \(i\)-th catalogued star. Equation \eqref{eq:los_approx} is the same first-order expansion used in Section \ref{subsec:combined}. Note that this expression can be re-normalized to impose unitary length. The measurement $\bm{y}$ corrupted by noise $\bm{\eta}$ at time \(t_k\) is then
\begin{equation}\label{eq:meas_model}
\bm y_k = \bm h(\bm x_k) + \bm\eta_k 
\end{equation}
where, using the first-order approximation, the measurement function reads
\begin{equation} \label{eq:measurement_function}
\bm h(\bm x_k^-) \approx \hat{\bm r}_i
+ \big(\bm I_3-\hat{\bm r}_i\hat{\bm r}_i^\top\big)\!\left(\frac{\bm v_k}{c}-\frac{\bm r_k}{r_i}\right)
\end{equation}
Note that, if desired, the measurement function can be re-normalized to achieve a unitary LoS direction. The measurement noise follows the small-angle assumption model (see~\ref{appendix:quest}) with covariance
\begin{equation}
\bm R_k = \sigma^2\big(\bm I_3 - \hat{\bm\rho}_i\hat{\bm\rho}_i^\top\big)
\end{equation}
where \(\sigma\) is the standard deviation of the angular measurement error and \(\hat{\bm\rho}_i\) denotes the predicted LoS direction at time $t_k$. Note that, if desired, an analyst can also include the star ephemerides error as
\begin{equation}
\bm R_k = [\sigma^2 + (\eta_i/\rho_i)^2]\big(\bm I_3 - \hat{\bm\rho}_i\hat{\bm\rho}_i^\top\big)
\end{equation}
However, the contribution due to star position uncertainty is much lower than the contribution to line-of-sight uncertainty, and, therefore, can be neglected. The Jacobian of \(\bm h\) with respect to \(\bm x\) is
\begin{equation}\label{eq:H_jacobian}
\bm H =
\frac{\partial \bm h}{\partial \bm x}
=
\begin{bmatrix}
-\dfrac{1}{r_i}\big(\bm I_3-\hat{\bm r}_i\hat{\bm r}_i^\top\big)
&\quad
\dfrac{1}{c}\big(\bm I_3-\hat{\bm r}_i\hat{\bm r}_i^\top\big)
\end{bmatrix}
\end{equation}
{Note that, if Eq.~\eqref{eq:measurement_function} is re-normalized, also the Jacobian has to consider the re-normalization. If this is preferred, the re-normalized Jacobian $\bm \tilde{H}$ reads} 
{\begin{equation}\label{eq:H_jacobian_normalized}
\bm \tilde{H} =
\frac{1}{\|\bm h(\bm{x}_k^{-})\|}
\left(
\bm I_3 - \hat{\bm\rho}_i \hat{\bm\rho}_i^\top
\right)
\begin{bmatrix}
-\dfrac{1}{r_i}\big(\bm I_3-\hat{\bm r}_i\hat{\bm r}_i^\top\big)
&\quad
\dfrac{1}{c}\big(\bm I_3-\hat{\bm r}_i\hat{\bm r}_i^\top\big)
\end{bmatrix}
\end{equation}}
Continuing considering the non-normalized formulation, at each measurement time \(t_k\), the discrete-time Kalman gain is
\begin{equation}
\bm K_k = \bm P_k^- \bm H^\top\big(\bm H\,\bm P_k^- \bm H^\top + \bm R_k\big)^{-1}
\end{equation}
and then, the state is updated as
\begin{equation}
\bm x_k^+ = \bm x_k^- + \bm K_k\big(\bm y_k - \bm h(\bm x_k^-)\big)
\end{equation}
and the covariance is updated using the Joseph form to preserve symmetry and numerical stability as
\begin{equation}\label{eq:joseph}
\bm P_k^+ = (\bm I - \bm K_k\bm H)\,\bm P_k^-\,(\bm I - \bm K_k\bm H)^\top
+ \bm K_k\,\bm R_k\,\bm K_k^\top
\end{equation}

\section{Performance} 
\label{sec:performances}

\begin{figure}[htbp]
  \centering
  \begin{subfigure}[b]{0.325\textwidth}
    \centering
    \includegraphics[width=\linewidth]{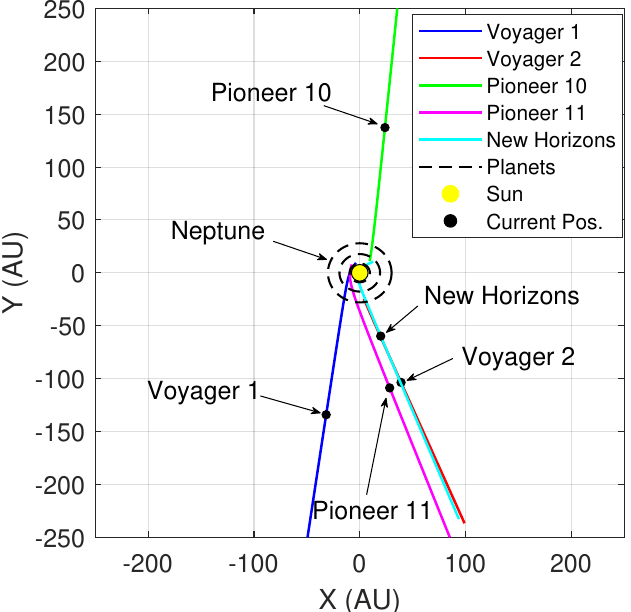}
    \caption{Projection on the XY plane.}
    \label{fig:trajXY}
  \end{subfigure}
  \hfill
  \begin{subfigure}[b]{0.325\textwidth}
    \centering
    \includegraphics[width=\linewidth]{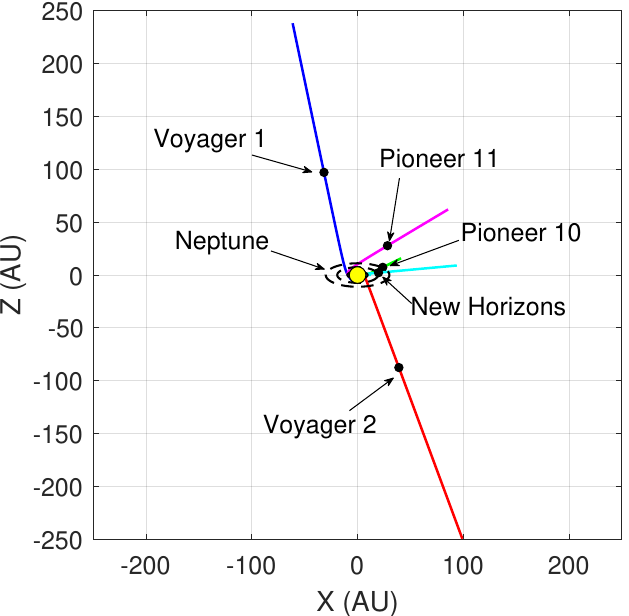}
    \caption{Projection on the XZ plane.}
    \label{fig:trajXZ}
  \end{subfigure}
  \hfill
  \begin{subfigure}[b]{0.325\textwidth}
    \centering
    \includegraphics[width=\linewidth]{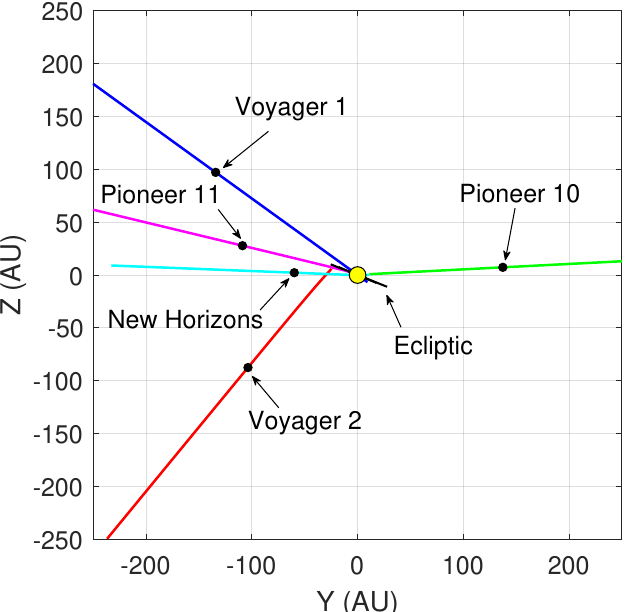}
    \caption{Projection on the YZ plane.}
    \label{fig:trajYZ}
  \end{subfigure}
  \caption{Heliocentric trajectories of the representative Voyager 1, Voyager 2, Pioneer 10, Pioneer 11, and New Horizons spacecraft in the J2000 reference frame. The Sun is denoted with a yellow circle, while the current spacecraft positions at the time of writing are indicated by black dots. Planetary orbits and the ecliptic plane are included for reference.}
  \label{fig:trajectories}
\end{figure}

Section~\ref{sec:methodology} and Section~\ref{sec:filter} have discussed methods for spacecraft position estimation through a least squares and Kalman filter formulations. This Section evaluates the performance of such methods within representative trajectories towards the outer region of the solar system. Performances are evaluated beyond 30 AU to the Sun, since the navigation approach relies on the apparent parallax measurements of nearby stars. To this aim, escape trajectories for spacecraft have been generated considering the asymptotic directions and velocities of Voyager 1 (VG-1), Voyager 2 (VG-2), Pioneer 10 (PR-10), Pioneer 11 (PR-11), and New Horizons (NH). This is also to evaluate navigation performances under representative cases of real spacecraft missions. Note that the sequences of gravity assists and maneuvering within the inner region of the solar system has been omitted, as the interest of this paper lies in evaluating the navigation performances when already beyond 30 AU from the Sun, which occurs after the last assist from planets. Therefore, the reference trajectories have been generated by back-propagating the current position, heading, and velocity conditions of these 5 space missions, which have been obtained from the JPL Horizons solar system data and ephemeris service. Ephemerides data in terms of position and velocity has been obtained at the time of writing. Note that trajectories have been plotted in the equatorial J2000 reference frame, as parameters in star catalogs are referred to this frame. Figure \ref{fig:trajectories} shows the generated representative trajectories, which constitute the numerical ground truth to evaluate performance of the two navigation methods. Note that the trajectories head towards different outward directions and follow different plane inclinations with respect to the ecliptic, thus leading to different parallax effects of nearby stars.

\subsection{Least Squares Estimation}

\begin{figure}[ht!]
  \centering
  \begin{subfigure}[b]{0.32\textwidth}
    \centering
    \includegraphics[width=\linewidth]{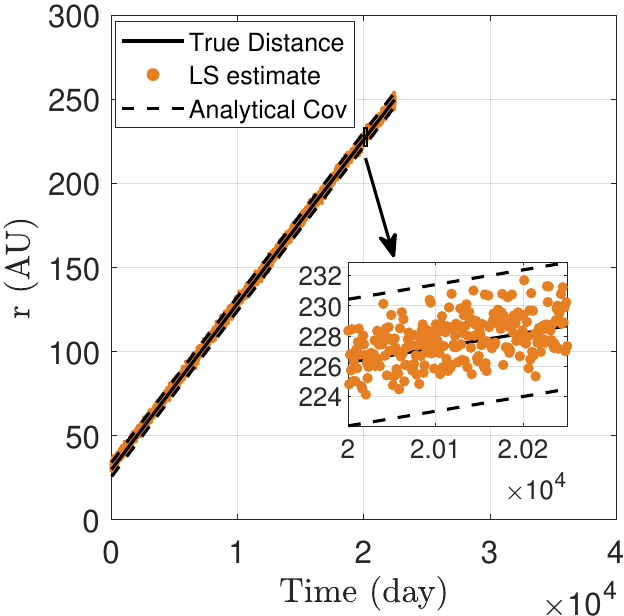}
    \caption{LS estimation for the VG-1 mission.}
    \label{fig:LS_vg1}
  \end{subfigure}
  \hfill
  \begin{subfigure}[b]{0.32\textwidth}
    \centering
    \includegraphics[width=\linewidth]{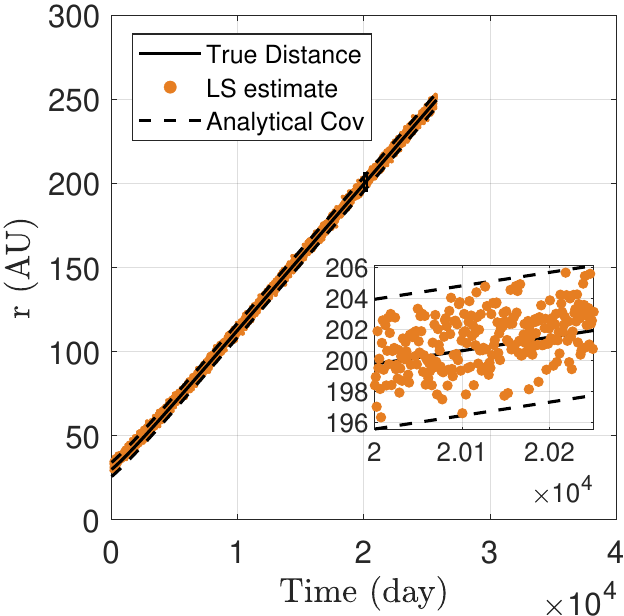}
    \caption{LS estimation for the VG-2 mission.}
    \label{fig:LS_vg2}
  \end{subfigure}
  \hfill
  \begin{subfigure}[b]{0.32\textwidth}
    \centering
    \includegraphics[width=\linewidth]{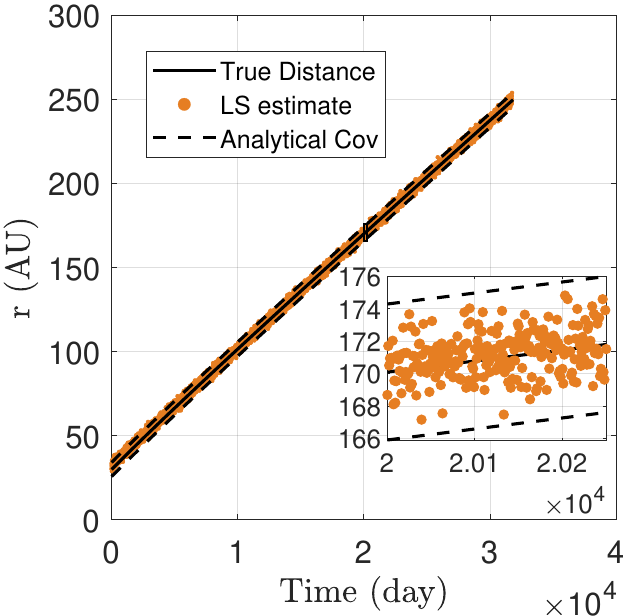}
    \caption{LS estimation for the PR-10 mission.}
    \label{fig:LS_pr10}
  \end{subfigure} \\[0.5cm]
    \begin{subfigure}[b]{0.32\textwidth}
    \centering
    \includegraphics[width=\linewidth]{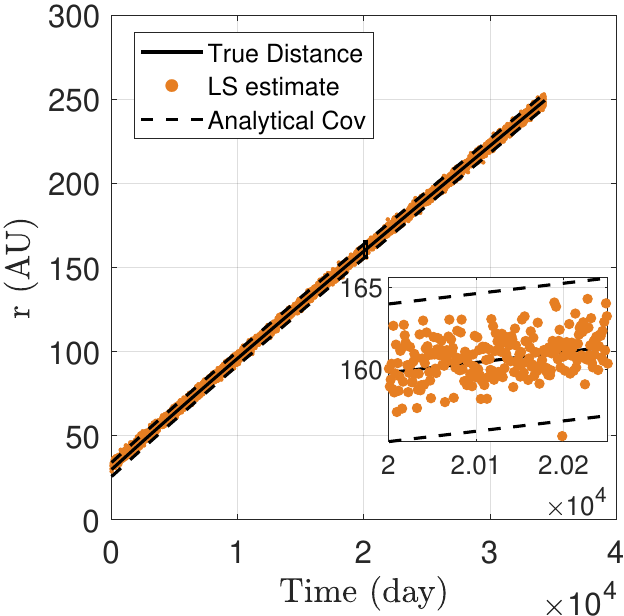}
    \caption{LS estimation for the PR-11 mission.}
    \label{fig:LS_pr11}
  \end{subfigure}
  \hspace{2cm}
  \begin{subfigure}[b]{0.32\textwidth}
    \centering
    \includegraphics[width=\linewidth]{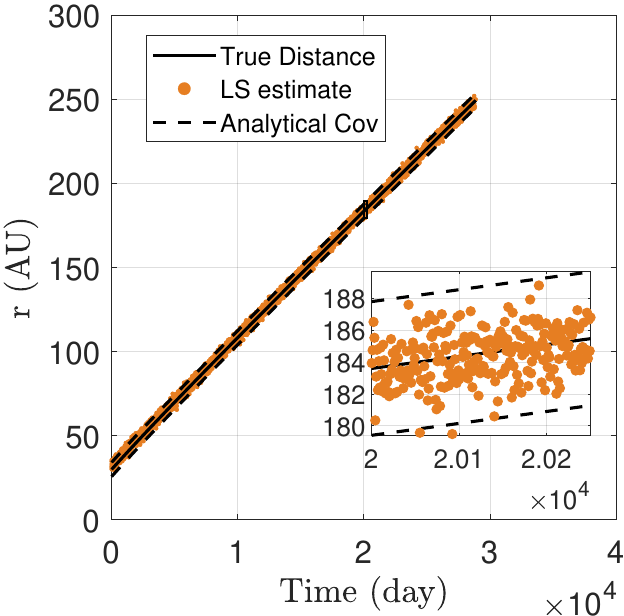}
    \caption{LS estimation for the NH mission.}
    \label{fig:LS_nh}
  \end{subfigure}
  \caption{Position estimation in a least squares sense along the (a) VG-1, (b) VG-2, (c) PR-10, (d) PR-11, and (e) NH mission trajectories. The estimation considers the parallax and aberration effects in the line-of-sight directions to nearby stars. The nearby stars are assumed to be tracked simultaneously along the mission trajectories.}
  \label{fig:LS_estimation}
\end{figure}

The performances of the least squares estimation method are reported in this section. The spacecraft position is first solved using the measured aberrated LoS directions to nearby stars as if they were non-aberrated, producing an initial position. In this way, the spacecraft coarse velocity can be estimated considering the last 10 position estimations, which in turn is used to correct aberration in the line-of-sight measurements and iterate on the spacecraft position. The de-aberrated line-of-sight directions are used only once, but in principle, an analyst is free to increase the number of iterations to increase the state estimation accuracy. It is assumed that 5 nearby stars are tracked simultaneously along the spacecraft mission trajectories with a 3$\sigma$ accuracy of 6 arcseconds, considering both attitude and image processing uncertainties. {The 5 stars per each mission are selected as the ones exhibiting the highest parallactic shift given the current position estimate. These are selected by a catalog of 14 nearby stars made up of the following stars: Lacaille (HIP 114046), Cygni A (HIP 104214), Cygni B (HIP 104217), Ross 154 (HIP 92403), Epsilon Eridani (HIP 16537), Sirius (HIP 32349), Procyon (HIP 37279), Proxima Centauri (HIP 70890), Alpha Centauri B (HIP 71681), Alpha Centauri A (HIP 71683), Barnard’s Star (HIP 87937), Ross 128 (HIP 57548), Struve (HIP 91768). Data about these stars are found in ~\ref{appendix:parallax}.} {Note that this method is used to show the suitability of using line-of-sight directions to nearby stars for position estimation, without seeking high accuracy for the results. The actual accuracy of star-based navigation is investigated in the Kalman filter in Section \ref{sec:KalmanFilterResults}, where the problem can be modeled in higher fidelity. Figure \ref{fig:LS_estimation} shows the spacecraft position estimates and the analytical covariance bounds computed according to Eq.~\ref{eq:least_squares_r} and Eq.~\ref{eq:cov_unweighted_sum}, respectively, showing agreement between the numerical solution and the covariance bounds.} {For all the trajectories, the analytical 3$\sigma$ covariance bounds of the least squares solution lie within 3.5 AU from the reference solution, while the numerical 3$\sigma$ covariance bounds lie within 2.5 AU from the reference solution. Note that there is no information about the spacecraft dynamics in the LS solution.}

\subsection{Kalman filter} \label{sec:KalmanFilterResults}
The extended Kalman filter for autonomous spacecraft navigation leverages the single line-of-sight measurements to nearby stars acquired one per time. All the simulations involving the Kalman filter consider a large initial uncertainty in the spacecraft position and velocity. In particular, the spacecraft position components start from an initial uncertainty sampled within 15 AU from the numerical truth, and the initial conditions for the velocity components are sampled within a $3\cdot10^{-4}$ AU/day 3$\sigma$ uncertainty with respect to the numerical truth. The propagation phase includes uncertainties in the spacecraft acceleration, which accounts for a $10^{-8}$ AU/day$^2$ random acceleration in all the components. The position of the stars has been propagated according to the measurement time, starting from the Hipparcos data catalog, and considering an isotropic uncertainty in the star location of 10 AU. The measurements in the filter account for stellar parallax and aberration using the first-order linearized approximation, as detailed in Section \ref{subsec:Update}. The noisy measurements are generated considering a 3$\sigma$ standard deviation of 6 arcseconds in the line-of-sight directions accounting for both attitude uncertainty and image processing errors. Noise is added to right ascension and declination angles, then unit vectors are reconstructed from these noisy angles. Every 7 days, the line-of-sight to just one star is acquired and given to the filter. This is to simulate a realistic scenario where just one star LoS is measured every seven days using a narrow FoV camera, which requires the spacecraft to slew and point towards a precise direction. {The star to be acquired is selected via a geometric observability score, which is function of the geometric parallax shift observed from the estimated spacecraft position. Note that, in order to maximize the geometric diversity among stars, a star that has already been tracked in the last two months is omitted, and the next best-scoring star for parallax is tracked. The selection algorithm is summarized in Algorithm~\ref{alg:StarSelection}.}

\begin{singlespace}
\begin{algorithm}[t]
\caption{{Geometric star selection based on estimated parallax observability}}
\label{alg:StarSelection}
\begin{algorithmic}[1]
\REQUIRE {Estimated spacecraft position $\bm{r}_k$, nearby stars catalog $\mathcal{S}$ with unit inertial directions $\hat{\bm{r}}_i$ and distances $r_i$, list of recently tracked stars $\mathcal{S}_{\text{recent}}$}
\ENSURE {Selected star $s^*$ for line-of-sight measurement}

\STATE {Compute estimated spacecraft radial direction:}
\[
\hat{\bm{r}}_{\text{k}} = \frac{\bm{r}_k}{\|\bm{r}_k\|}
\]

\FOR{{each star $s_i \in \mathcal{S}$}}
    \STATE {Compute angular separation between spacecraft radial direction and star direction:}
    \[
    \cos\phi_i = \hat{\bm{r}}_i^\top \hat{\bm{r}}_{\text{k}}
    \]
    \STATE {Evaluate geometric parallax factor:}
    \[
    \sin\phi_i = \sqrt{1 - \cos^2\phi_i}
    \]
    \STATE {Compute geometric observability score:}
    \[
    J_i = \frac{\sin\phi_i}{r_i}
    \]
\ENDFOR

\STATE {Exclude recently tracked stars to promote measurement diversity:}
\[
\mathcal{S}_{\text{cand}} = \mathcal{S} \setminus \mathcal{S}_{\text{recent}}
\]

\STATE {Select the star maximizing the geometric score:}
\[
s^* = \arg\max_{s_i \in \mathcal{S}_{\text{cand}}} J_i
\]

\RETURN {$s^*$}
\end{algorithmic}
\end{algorithm}
\end{singlespace}

\begin{figure}[ht!]
  \centering
  \begin{subfigure}[b]{0.325\textwidth}
    \centering
    \includegraphics[width=\linewidth]{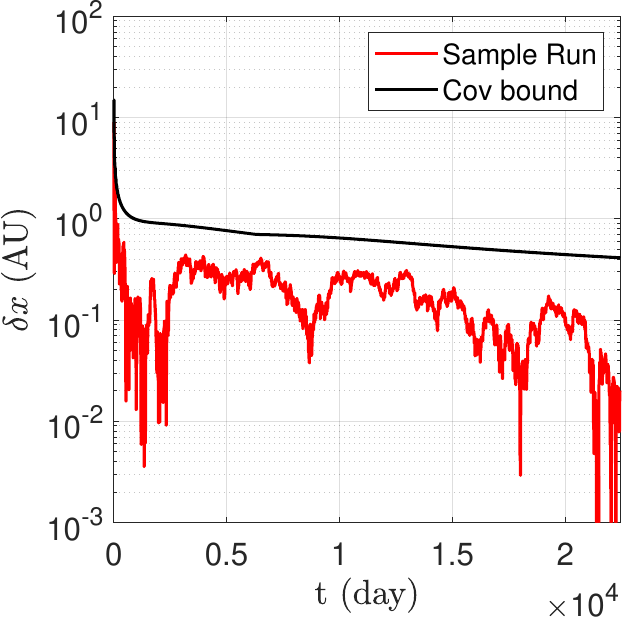}
    \caption{$\delta x$ sample run error and 3$\sigma$ bounds.}
    \label{fig:sub_x}
  \end{subfigure}
  \hfill
  \begin{subfigure}[b]{0.325\textwidth}
    \centering
    \includegraphics[width=\linewidth]{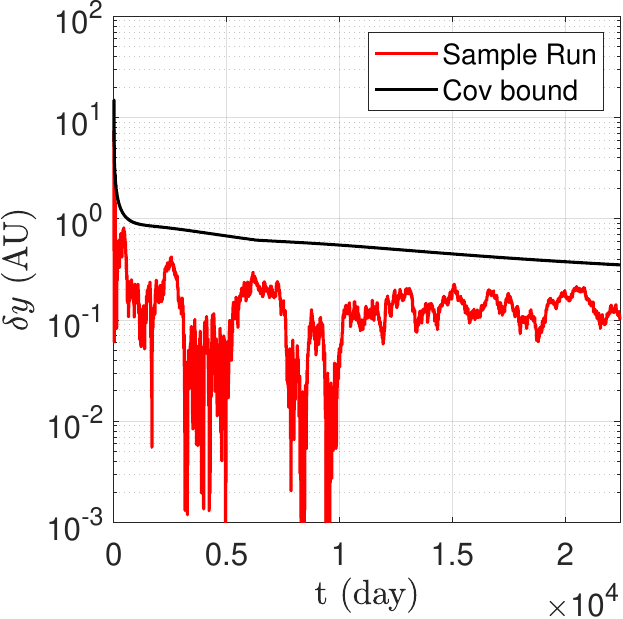}
    \caption{$\delta y$ sample run error and 3$\sigma$ bounds.}
    \label{fig:sub_y}
  \end{subfigure}
  \hfill
  \begin{subfigure}[b]{0.325\textwidth}
    \centering
    \includegraphics[width=\linewidth]{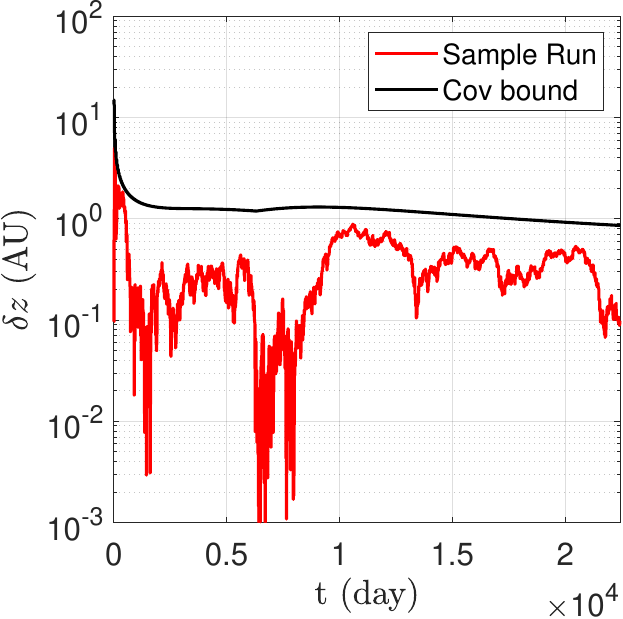}
    \caption{$\delta z$ sample run error and 3$\sigma$ bounds.}
    \label{fig:sub_z}
  \end{subfigure}
    \begin{subfigure}[b]{0.325\textwidth}
    \centering
    \includegraphics[width=\linewidth]{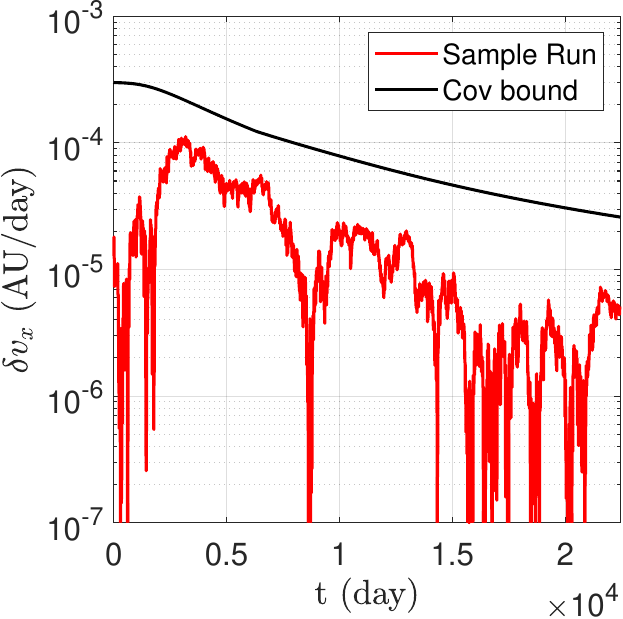}
    \caption{$\delta v_x$ sample run error and 3$\sigma$ bounds.}
    \label{fig:sub_vx}
  \end{subfigure}
  \hfill
  \begin{subfigure}[b]{0.325\textwidth}
    \centering
    \includegraphics[width=\linewidth]{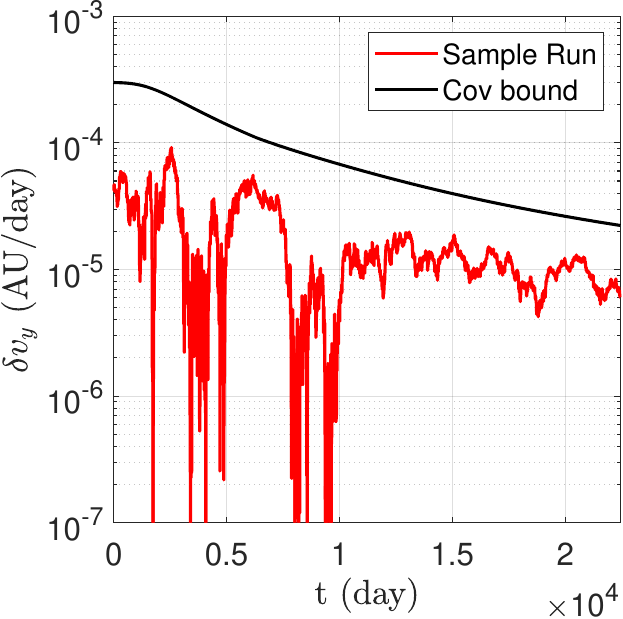}
    \caption{$\delta v_y$ sample run error and 3$\sigma$ bounds.}
    \label{fig:sub_vy}
  \end{subfigure}
  \hfill
  \begin{subfigure}[b]{0.325\textwidth}
    \centering
    \includegraphics[width=\linewidth]{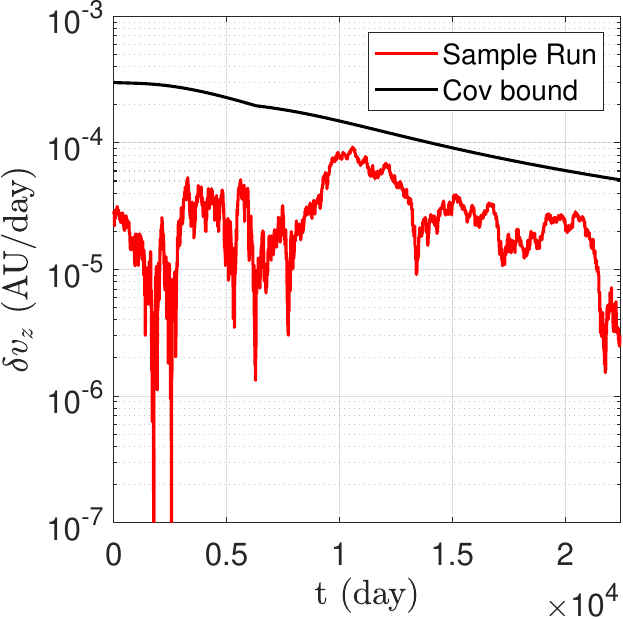}
    \caption{$\delta v_z$ sample run error and 3$\sigma$ bounds.}
    \label{fig:sub_vz}
  \end{subfigure}
  \caption{Kalman filter outputs in terms of sample run error and filter covariance bounds for position error ($\delta x$, $\delta y$, $\delta z$) and velocity error ($\delta v_x$, $\delta v_y$, $\delta v_z$) components for the VG1 mission trajectory in the J2000 reference frame. }
  \label{fig:samplekalmanfilter}
\end{figure}

{Figure \ref{fig:samplekalmanfilter} shows a sample run of the Kalman filter including the sample errors and the filter-estimated covariance bounds for the spacecraft position ($\delta x$, $\delta y$, $\delta z$) and velocity ($\delta v_x$, $\delta v_y$, $\delta v_z$) components in the J2000 frame. This simulation is along the Voyager 1 trajectory, where the spacecraft autonomously acquires one line-of-sight direction to a single star every 7 days, starting from 30 AU from the Sun up to 250 AU. We can see that, with this conservative assumption, the velocity errors and covariance bounds are decreasing in time. The same applies to the position error and covariance plots, but with a slower convergence rate. We can note that the filter predicted 3$\sigma$ covariance bounds are kept well within 1 AU of uncertainty in terms of position even at 250 AU from the Sun, and within $4\cdot10^{-5}$ AU/day uncertainty in terms of velocity. Note that these values correspond to relative errors of 0.4$\%$ in terms of both position and velocity with respect to the numerical ground truth values.}

\clearpage

\begin{figure}[htbp]
  \centering
  \begin{subfigure}[b]{0.245\textwidth}
    \centering
    \includegraphics[width=\linewidth]{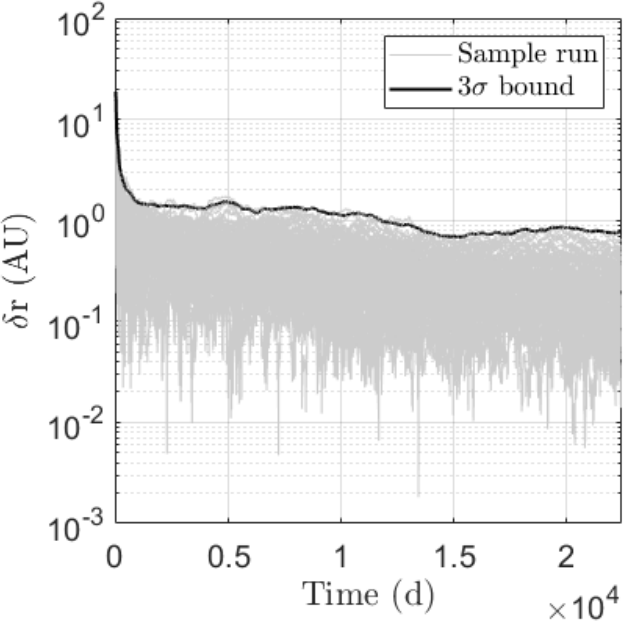}
    \caption{MonteCarlo samples and 3$\sigma$ bounds in position error for the VG-1 mission trajectory.}
    \label{fig:r_vg1}
  \end{subfigure}
  \begin{subfigure}[b]{0.245\textwidth}
    \centering
    \includegraphics[width=\linewidth]{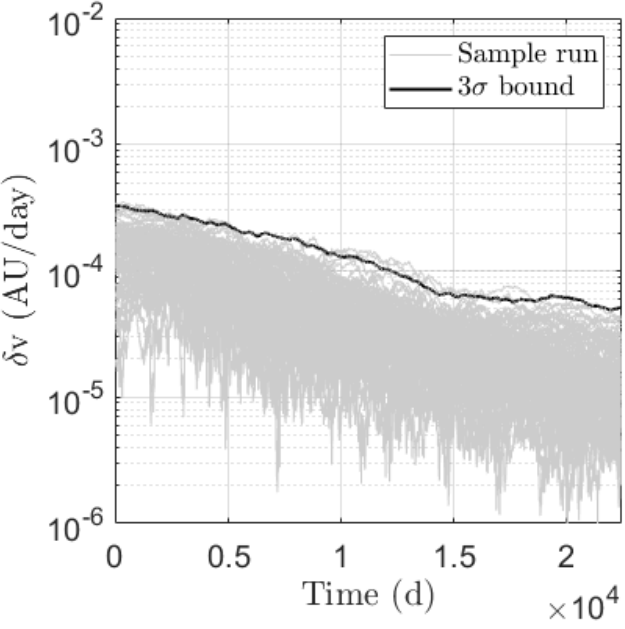}
    \caption{MonteCarlo samples and 3$\sigma$ bounds in velocity error for the VG-1 mission trajectory.}
    \label{fig:v_vg1}
  \end{subfigure}
  \begin{subfigure}[b]{0.245\textwidth}
    \centering
    \includegraphics[width=\linewidth]{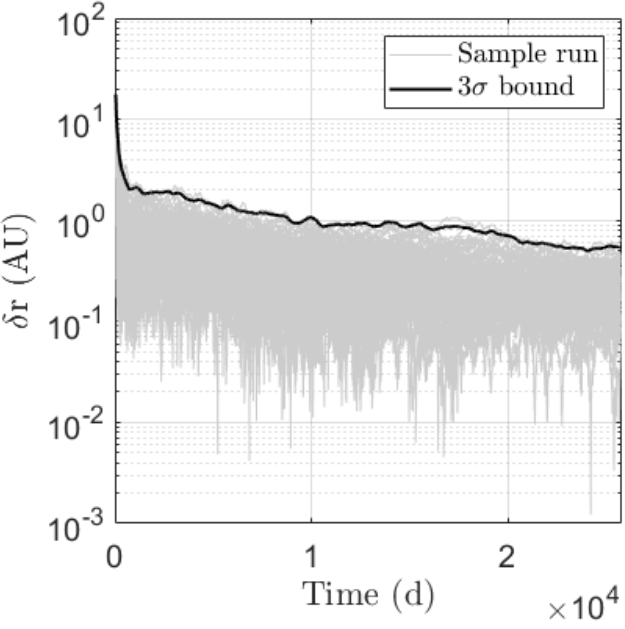}
    \caption{MonteCarlo samples and 3$\sigma$ bounds in position error for the VG-2 mission trajectory.}
    \label{fig:r_vg2}
  \end{subfigure} 
    \begin{subfigure}[b]{0.245\textwidth}
    \centering
    \includegraphics[width=\linewidth]{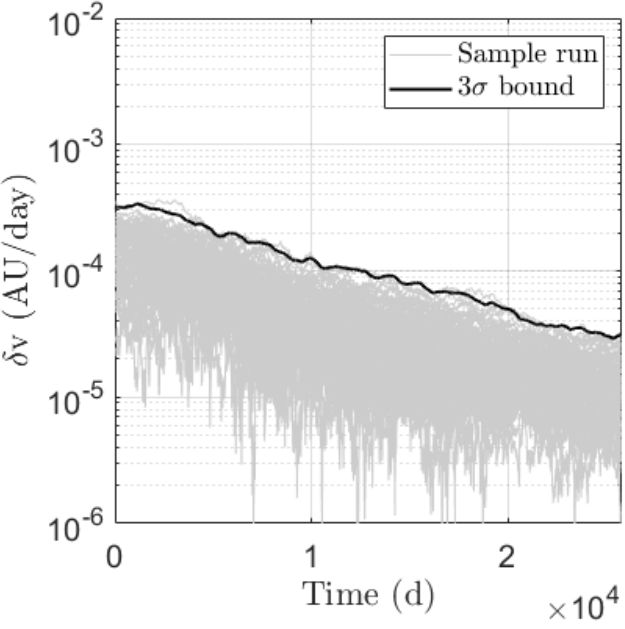}
    \caption{MonteCarlo samples and 3$\sigma$ bounds in velocity error for the VG-2 mission trajectory.}
    \label{fig:v_vg2}
  \end{subfigure} \\[0.25cm]
  \begin{subfigure}[b]{0.245\textwidth}
    \centering
    \includegraphics[width=\linewidth]{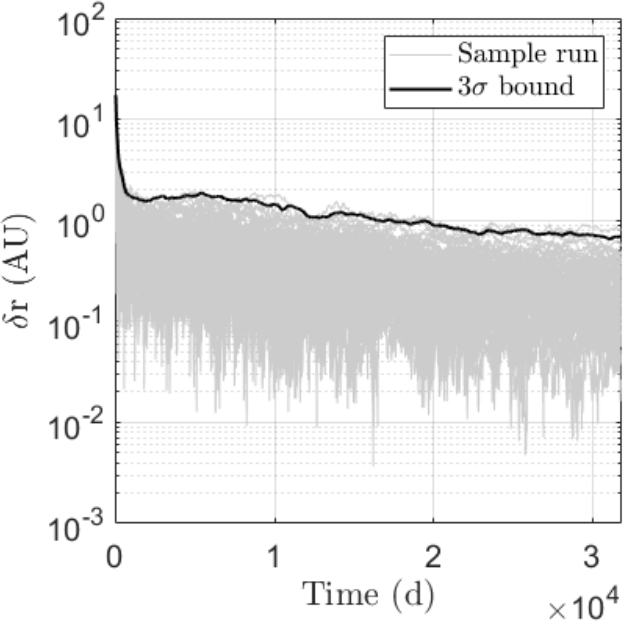}
    \caption{MonteCarlo samples and 3$\sigma$ bounds in position error for the PR-10 mission trajectory.}
    \label{fig:r_pr10}
  \end{subfigure}
  \begin{subfigure}[b]{0.245\textwidth}
    \centering
    \includegraphics[width=\linewidth]{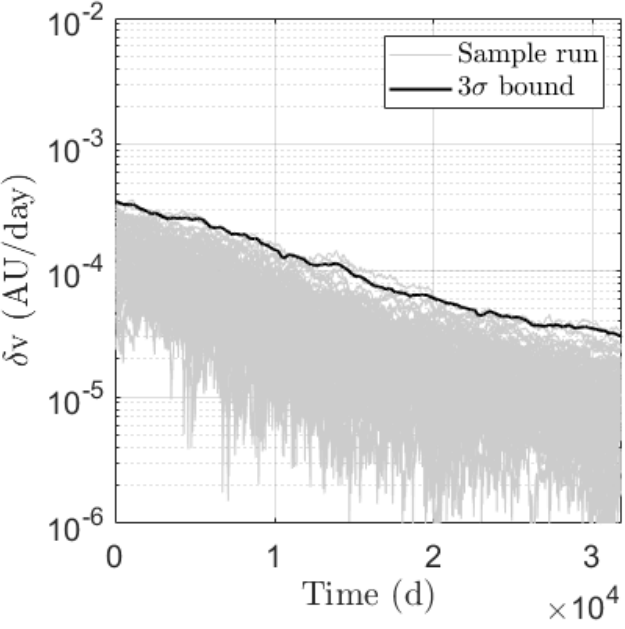}
    \caption{MonteCarlo samples and 3$\sigma$ bounds in velocity error for the PR-10 mission trajectory.}
    \label{fig:v_pr10}
  \end{subfigure}
    \begin{subfigure}[b]{0.245\textwidth}
    \centering
    \includegraphics[width=\linewidth]{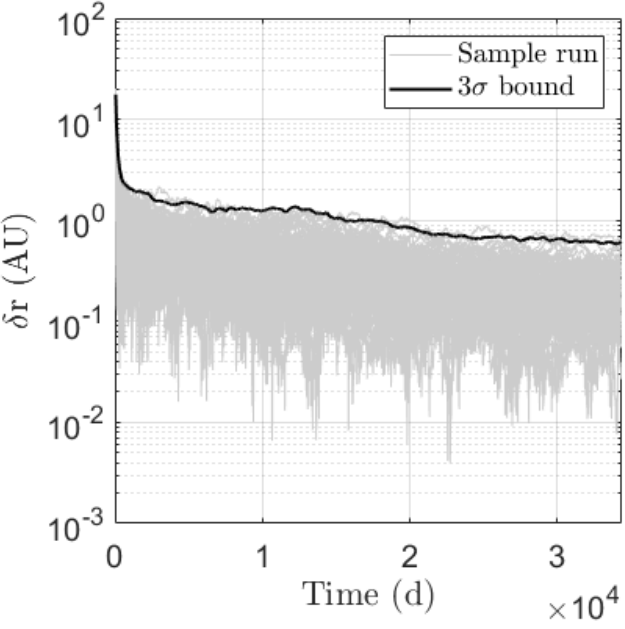}
    \caption{MonteCarlo samples and 3$\sigma$ bounds in position error for the PR-11 mission trajectory.}
    \label{fig:r_pr11}
  \end{subfigure}
  \begin{subfigure}[b]{0.245\textwidth}
    \centering
    \includegraphics[width=\linewidth]{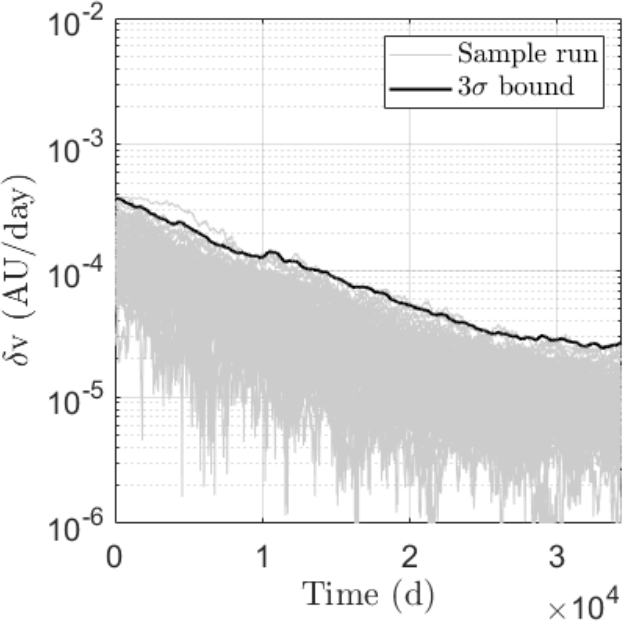}
    \caption{MonteCarlo samples and 3$\sigma$ bounds in velocity error for the PR-11 mission trajectory.}
    \label{fig:v_pr11}
  \end{subfigure} \\[0.25cm]
    \begin{subfigure}[b]{0.245\textwidth}
    \centering
    \includegraphics[width=\linewidth]{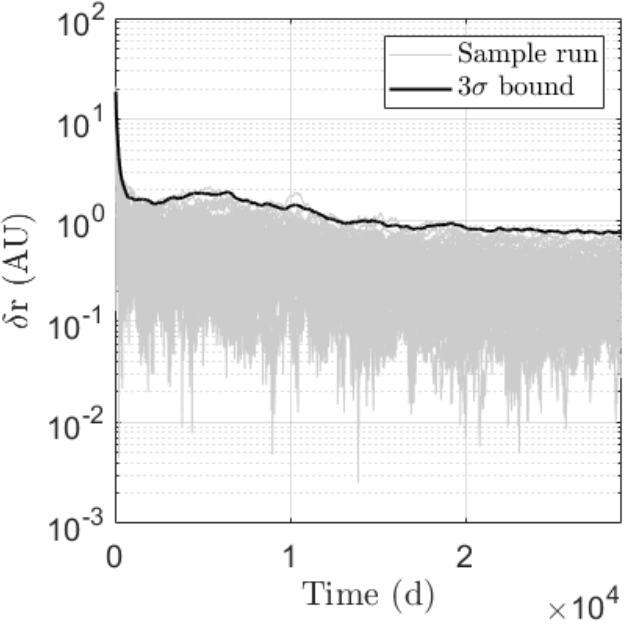}
    \caption{MonteCarlo samples and 3$\sigma$ bounds in position error for the NH mission trajectory.}
    \label{fig:r_nh}
  \end{subfigure}
  \begin{subfigure}[b]{0.245\textwidth}
    \centering
    \includegraphics[width=\linewidth]{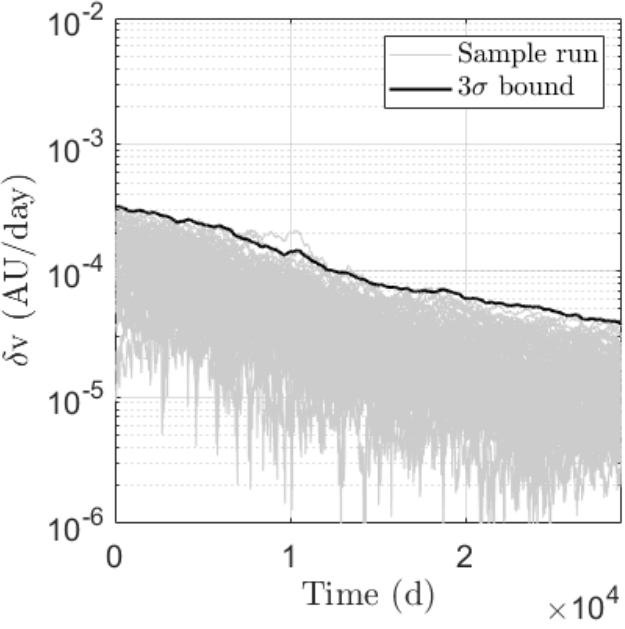}
    \caption{MonteCarlo samples and 3$\sigma$ bounds in velocity error for the NH mission trajectory.}
    \label{fig:v_nh}
  \end{subfigure}
  \caption{Sample errors and numerical covariance bounds across Monte carlo runs for VG-1, VG-2, PR-10, PR-11, and NH trajectories. Results shown in terms of position and velocity norm errors and covariance bounds in the J2000 reference frame. A total of 1000 samples have been run per mission.}
  \label{fig:MonteCarlo}
\end{figure}

\begin{figure}[htbp]
  \centering
  \begin{subfigure}[b]{0.32\textwidth}
    \centering
    \includegraphics[width=\linewidth]{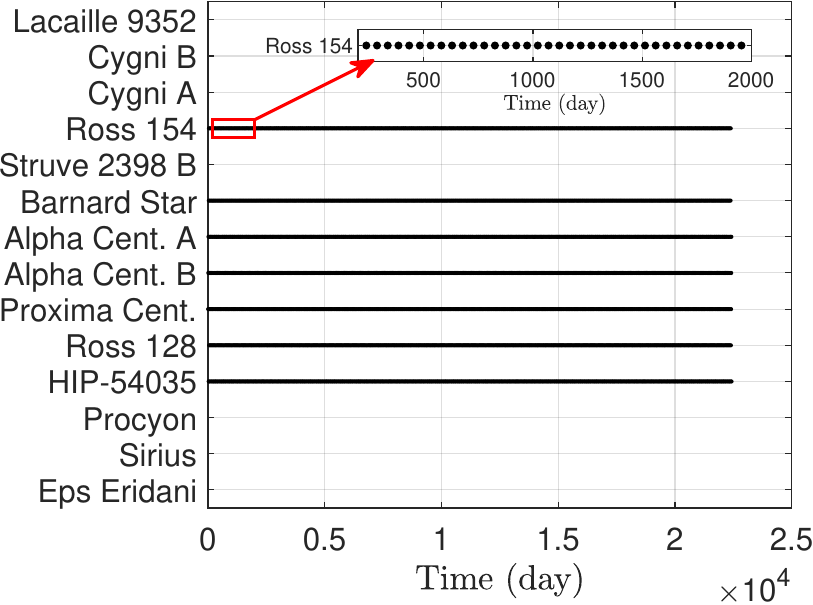}
    \caption{Stars tracked along the VG-1 representative mission trajectory.}
    \label{fig:stars_vg1}
  \end{subfigure}
  \hfill
  \begin{subfigure}[b]{0.32\textwidth}
    \centering
    \includegraphics[width=\linewidth]{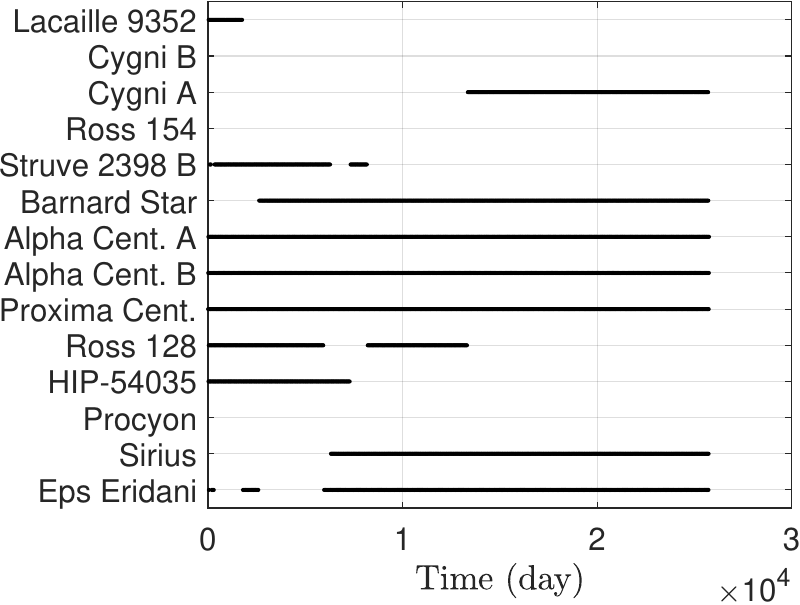}
    \caption{Stars tracked along the VG-2 representative mission trajectory.}
    \label{fig:stars_vg2}
  \end{subfigure}
  \hfill
  \begin{subfigure}[b]{0.32\textwidth}
    \centering
    \includegraphics[width=\linewidth]{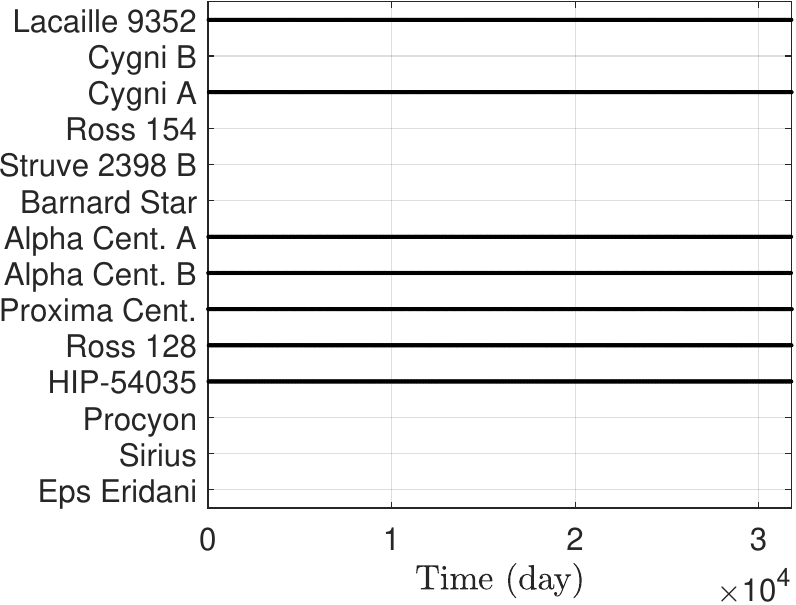}
    \caption{Stars tracked along the PR-10 representative mission trajectory.}
    \label{fig:stars_pr10}
  \end{subfigure} \\[0.5cm]
    \begin{subfigure}[b]{0.32\textwidth}
    \centering
    \includegraphics[width=\linewidth]{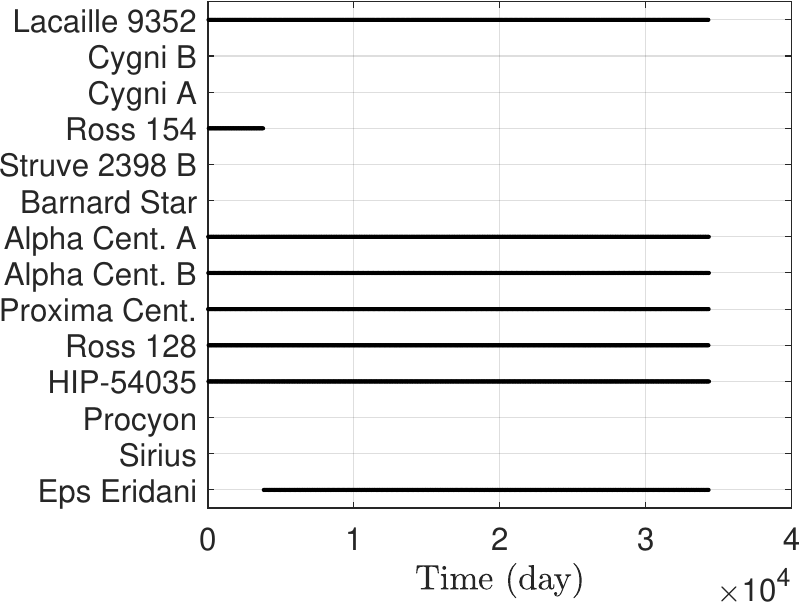}
    \caption{Stars tracked along the PR-11 representative mission trajectory.}
    \label{fig:stars_pr11}
  \end{subfigure}
  \hspace{2cm}
  \begin{subfigure}[b]{0.32\textwidth}
    \centering
    \includegraphics[width=\linewidth]{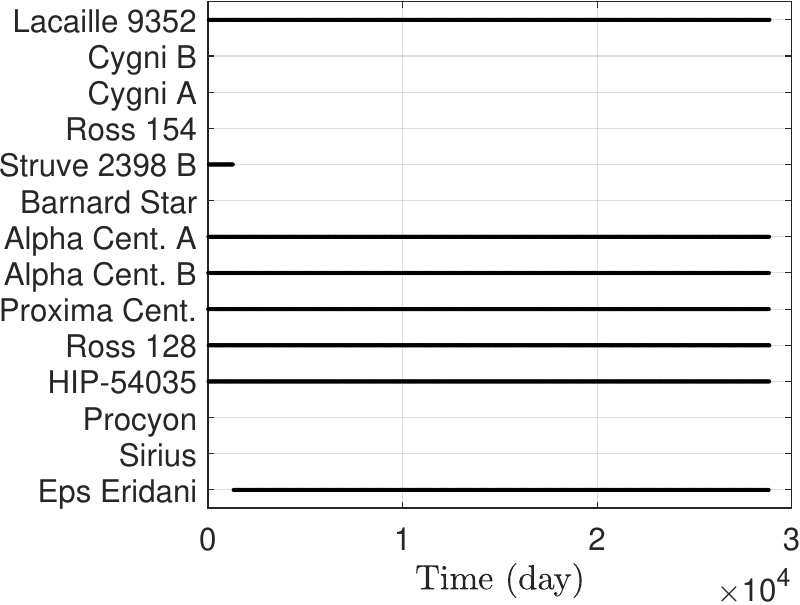}
    \caption{Stars tracked along the NH representative mission trajectory.}
    \label{fig:stars_nh}
  \end{subfigure}
  \caption{Stars tracked along the (a) VG-1, (b) VG-2, (c) PR-10, (d) PR-11, and (e) NH mission trajectories. {The Hipparcos IDs of the tracked stars are as follows: Lacaille: HIP 114046, Cygni A: HIP 104214, Cygni B: HIP 104217, Ross 154: HIP 92403, Epsilon Eridani: HIP 16537, Sirius: HIP 32349, Procyon: HIP 37279, Proxima Centauri: HIP 70890, Alpha Centauri B: HIP 71681, Alpha Centauri A: HIP 71683, Barnard’s Star: HIP 87937, Ross 128: HIP 57548, Struve: HIP 91768)}.}
  \label{fig:stars_tracked}
\end{figure}

{Figure \ref{fig:MonteCarlo} shows Monte Carlo simulations running 1000 Kalman filter samples for each of the 5 spacecraft trajectories. Results are shown in terms of position norm error ($\delta r$) and velocity norm error ($\delta v$). Both the sample errors and the computed 3$\sigma$ bounds across the samples are plotted. We can note that, regardless of the spacecraft mission trajectory, the filter is able to provide accuracies better than 1 AU in terms of position and better than $4\cdot10^{-5}$ AU/day in terms of velocity, converging from an initial large uncertainties of 15 AU in terms of position and 0.5$\cdot10^{-3}$ AU/day in terms of velocity. Figure \ref{fig:stars_tracked} shows the stars tracked along the VG-1, VG-2, PR-10, PR-11, and NH mission trajectories. Each star is tracked in a sequential way according to Algorithm~\ref{alg:StarSelection}, and just one line-of-sight measurement to a single star is acquired per week. Each star is tracked for a maximum of once every two months to allow for geometric diversity and realistic mission operations scenario, as shown in the zoom of the first subplot (See Fig.~\ref{fig:stars_vg1}). Note that the optimal stars to be tracked evolve as the spacecraft moves towards the outer region of the solar system.}

\clearpage

\clearpage

\begin{figure}[htbp]
  \centering
  \begin{subfigure}[b]{0.325\textwidth}
    \centering
    \includegraphics[width=\linewidth]{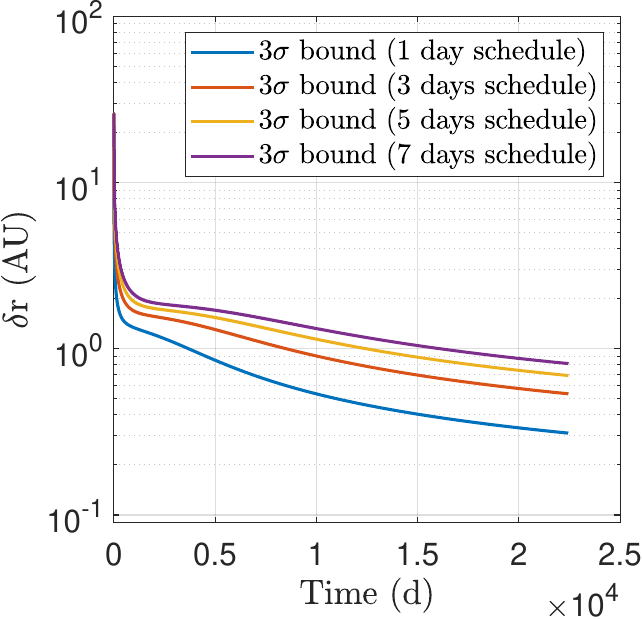}
    \caption{{Effect of acquisition schedule on position estimation (3$\sigma$ bounds).}}
    \label{fig:schedule_dr}
  \end{subfigure}
  \hspace{1cm}
  \begin{subfigure}[b]{0.325\textwidth}
    \centering
    \includegraphics[width=\linewidth]{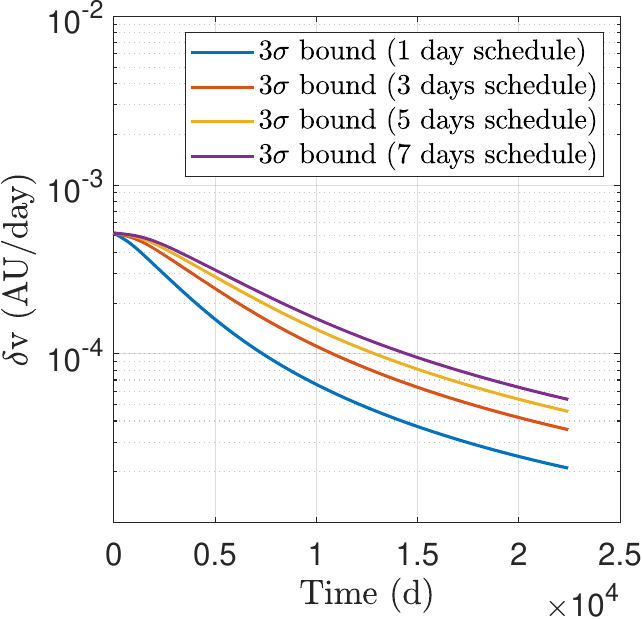}
    \caption{{Effect of acquisition schedule on velocity estimation (3$\sigma$ bounds).}}
    \label{fig:schedule_dv}
  \end{subfigure}
  \caption{{Sensitivity of the navigation performances (3$\sigma$ filter covariance bounds in position and velocity) to the measurements schedule. Results are shown considering acquisition of one star measurement taken each day, each three days, each five days, and each seven days for the representative VG-1 mission trajectory.}}
  \label{fig:schedule}
\end{figure}

{Eventually, the sensitivity of the navigation performances to the measurements schedule is shown in Figure~\ref{fig:schedule}. Results are shown considering an acquisition frequency of one star measurement taken each day, each three days, each five days, and each seven days, respectively, for the representative VG-1 mission trajectory. It is possible to note how an increased measurement frequency leads to superior performances of the trajectory estimation in both position and velocity components. As an example, considering one star acquired every day, the filter determines the spacecraft position within a 0.3 AU error in 3$\sigma$ confidence, and the velocity within 2$\cdot 10^{-4}$ AU/day in 3$\sigma$ confidence at a distance of 250 AU to the Sun.}   

\section{Conclusion} \label{sec:conclusions}

This work investigated the feasibility of using the stellar parallactic shift as an autonomous navigation source for spacecraft operating in the outer regions of the solar system. A complete star line-of-sight observation model accounting for parallax and aberration effects was developed. This model was incorporated into both a least squares estimation method and a sequential Kalman filter estimation framework. The formulation enables the spacecraft to infer its heliocentric position and velocity from line-of-sight measurements to stars in proximity to the solar system. This approach is applicable to both small and large spacecraft architectures and is particularly relevant for missions with constrained communication opportunities or long cruise phases. The results show that, under realistic angular measurement errors and observation cadences, the navigation filter converges even from large initial uncertainties. At distances up to 250 AU, the method achieves sub-AU position accuracy and velocity estimation on the order of $10^{-5}$ AU/day. These findings indicate that, while distant stars can still be employed for attitude reconstruction, nearby stars can provide meaningful information for autonomous position and velocity estimation at deep-space distances where conventional radiometric tracking becomes a limiting factor.

\section*{Funding Sources}
This work received no external source of funding.

\section*{Acknowledgments}
This work received no external source of funding.

\clearpage

\appendices

\makeatletter
\renewcommand\thesection{Appendix~\Alph{section}}
\makeatother

\section{Transformation from star catalogue data to Cartesian state vectors} 
\label{appendix:catalogue_transform}

Astrometric catalogues such as Hipparcos or Gaia provide the position and motion of stars in terms of right ascension ($\alpha$), declination ($\delta$), parallax ($p$), proper motion in right ascension ($\mu_{\alpha}$), proper motion in declination ($\mu_{\delta}$), and radial velocity ($v_\textrm{r}$) in the ICRF. These quantities fully describe the kinematics of a star and can be converted into Cartesian position ($\bm{r}$) and velocity vectors ($\bm{v}$) as follows. Note that the catalogue parameters are given in a heliocentric spherical coordinate system. Therefore, for each star, the orthonormal triad 
\([\bm{\hat{r}}_\textrm{s}, \bm{\hat{\alpha}}_\textrm{s}, \bm{\hat{\delta}}_\textrm{s}]\) 
can be constructed from its right ascension and declination as 
\begin{equation}
\bm{\hat{r}}_\textrm{s} =
{\renewcommand{\arraystretch}{0.8}\begin{bmatrix}
\cos \delta \cos \alpha \\
\cos \delta \sin \alpha \\
\sin \delta
\end{bmatrix}}; \qquad
\bm{\hat{\alpha}}_\textrm{s} =
{\renewcommand{\arraystretch}{0.8}\begin{bmatrix}
- \sin \alpha \\
\cos \alpha \\
0
\end{bmatrix}}; \qquad
\bm{\hat{\delta}}_\textrm{s} =
{\renewcommand{\arraystretch}{0.8}\begin{bmatrix}
- \sin \delta \cos \alpha \\
- \sin \delta \sin \alpha \\
\cos \delta
\end{bmatrix}}
\end{equation}
\subsection*{Position vector of a star}
The stellar parallax is the apparent angular displacement of a star on the celestial sphere caused by the change in the observer's point of view due to the Earth's orbital motion. By definition, a star with a parallax of $1$ arcsecond lies at a distance of $1$ parsec. Therefore, the heliocentric distance of a star $r_\textrm{s}$ can be obtained from the catalogued parallax as $r_\textrm{s}$ = $p_\textrm{s}^{-1}$, where $p_\textrm{s}$ is the stellar parallax in arcseconds and $r_\textrm{s}$ is expressed in parsec. Note that $r_\textrm{s}$ can be expressed in AU through the conversion factor 1 pc = $206{,}264.8$ AU. The heliocentric position vector of the star is then
\begin{equation} \label{eq:stellar_pos}
\bm{r}_\textrm{s} = r_\textrm{s} \, \bm{\hat{r}}_\textrm{s}
\end{equation}
\subsection*{Velocity vector of a star}
Following the spherical coordinates, the proper motion in right ascension is  $\mu_{\alpha} = \dot{\alpha} \cos \delta$, while in declination it is $\mu_\delta = \dot{\delta}$. Note that the proper motion in right ascension can be found in the catalogs as $\mu_{\alpha*}$ instead of $\mu_\alpha$ to emphasize the inclusion of the $\cos\delta$ factor. Let us now note that the velocity vector in Cartesian coordinates can be expressed in its components as
\begin{equation} \label{eq:stellar_vel}
\bm{v} = v_\textrm{r} \, \bm{\hat{r}}_\textrm{s} + v_\alpha \, \bm{\hat{\alpha}}_\textrm{s} + v_\delta \, \bm{\hat{\delta}}_\textrm{s}
\end{equation}
While $v_r$ is directly retrievable from the catalogs, the other components can be determined as 
\begin{equation} \label{eq:velocity_components}
v_\alpha = r_\textrm{s} \, \mu_{\alpha} \quad ; \quad
v_\delta = r_\textrm{s} \, \mu_\delta
\end{equation}
Note that, while the radial velocity is given in km/s, the proper motion parameters in the catalogues are expressed in arcsec/year. Therefore, they have to be converted into rad/s before being used in Eq.~\eqref{eq:velocity_components} for consistency, and then, all the units of measure of the length have to be uniformly expressed in km or in AU. The conversion factor from AU to km is 1 AU = 1.4959787 $\cdot$ $10^8$ km. The conversion factor from arcsec/year to rad/s is 1 arcsec/year = 1.5373 $\cdot$ 10$^{-13}$ rad/s.

\clearpage

\section{Parallax of nearby stars for different baselines} \label{appendix:parallax}

\begin{table}[h!]
\centering
\caption{Nearby stars within 15 light years to the SSB. Data retrieved and processed from the Hipparcos catalog~\cite{perryman1997hipparcos}. ID: Hipparcos identification number, Mag: stellar magnitude, RA: right ascension, DE: declination, Parallax: stellar parallax in milliarcseconds, pmRA: proper motion in right ascension, pmDE: proper motion in declination, Dist: distance in light years, $\Delta \theta_{75}$, $\Delta \theta_{150}$, $\Delta \theta_{250}$: stellar apparent angular shifts for 75 AU, 150 AU, and 250 AU transversal baselines. All astrometric quantities are referred to epoch J1991.25.}
\begin{tabular}{ccccccccccc} 
\hline \hline
ID & Mag & RA & DE & Parallax & pmRA & pmDE & Dist & $\Delta \theta_{75}$ & $\Delta \theta_{150}$ & $\Delta \theta_{250}$ \\ 
& - & deg & deg & mas & mas/yr & mas/yr & ly & arcsec & arcsec & arcsec \\ \hline
70890 & 11.01 & 217.4489 & -62.6814 & 772.330 & -3775.64 & 768.16 & 4.22 & 57.92 & 115.84 & 193.08 \\
71681 & 1.35 & 219.9141 & -60.8395 & 742.120 & -3600.35 & 952.11 & 4.39 & 55.66 & 111.32 & 185.53 \\
71683 & -0.01 & 219.9204 & -60.8351 & 742.120 & -3678.19 & 481.84 & 4.39 & 55.66 & 111.32 & 185.53 \\
87937 & 9.54 & 269.4540 & 4.6683 & 549.010 & -797.84 & 10326.93 & 5.94 & 41.18 & 82.35 & 137.25 \\
54035 & 7.49 & 165.8359 & 35.9815 & 392.400 & -580.20 & -4767.09 & 8.31 & 29.43 & 58.86 & 98.10 \\
32349 & -1.44 & 101.2885 & -16.7131 & 379.210 & -546.01 & -1223.08 & 8.60 & 28.44 & 56.88 & 94.80 \\
92403 & 10.37 & 282.4540 & -23.8358 & 336.480 & 637.55 & -192.47 & 9.69 & 25.24 & 50.47 & 84.12 \\
16537 & 3.72 & 53.2351 & -9.4583 & 310.750 & -976.44 & 17.97 & 10.50 & 23.31 & 46.62 & 77.69 \\
114046 & 7.35 & 346.4465 & -35.8563 & 303.900 & 6767.26 & 1326.66 & 10.73 & 22.79 & 45.58 & 75.98 \\
57548 & 11.12 & 176.9335 & 0.8075 & 299.580 & 605.62 & -1219.23 & 10.89 & 22.47 & 44.94 & 74.90 \\
104214 & 5.20 & 316.7118 & 38.7415 & 287.130 & 4155.10 & 3258.90 & 11.36 & 21.53 & 43.06 & 71.78 \\
37279 & 0.40 & 114.8272 & 5.2275 & 285.930 & -716.57 & -1034.58 & 11.41 & 21.44 & 42.88 & 71.48 \\
104217 & 6.05 & 316.7175 & 38.7344 & 285.420 & 4107.40 & 3143.72 & 11.43 & 21.41 & 42.81 & 71.36 \\
91772 & 9.70 & 280.7021 & 59.6224 & 284.480 & -1393.20 & 1845.73 & 11.46 & 21.34 & 42.69 & 71.12 \\
91768 & 8.94 & 280.7009 & 59.6260 & 280.280 & -1326.88 & 1802.12 & 11.64 & 21.02 & 42.03 & 70.07 \\
1475 & 8.09 & 4.5856 & 44.0220 & 280.270 & 2888.92 & 410.58 & 11.64 & 21.02 & 42.03 & 70.07 \\
108870 & 4.69 & 330.8227 & -56.7798 & 275.760 & 3959.97 & -2538.84 & 11.83 & 20.69 & 41.38 & 68.94 \\
8102 & 3.49 & 26.0214 & -15.9396 & 274.170 & -1721.82 & 854.07 & 11.90 & 20.57 & 41.14 & 68.54 \\
5643 & 12.10 & 18.1246 & -17.0005 & 269.050 & 1210.09 & 646.95 & 12.12 & 20.18 & 40.36 & 67.26 \\
36208 & 9.84 & 111.8507 & 5.2348 & 263.260 & 571.27 & -3694.25 & 12.39 & 19.75 & 39.50 & 65.82 \\
24186 & 8.86 & 77.8967 & -45.0045 & 255.260 & 6506.05 & -5731.39 & 12.78 & 19.15 & 38.30 & 63.82 \\
105090 & 6.69 & 319.3238 & -38.8646 & 253.370 & -3259.00 & -1146.99 & 12.87 & 19.01 & 38.02 & 63.34 \\
110893 & 9.59 & 337.0017 & 57.6970 & 249.520 & -870.23 & -471.10 & 13.07 & 18.71 & 37.42 & 62.38 \\
30920 & 11.12 & 97.3458 & -2.8125 & 242.890 & 694.73 & -618.62 & 13.43 & 18.22 & 36.44 & 60.72 \\
72511 & 11.72 & 222.3896 & -26.1060 & 235.240 & -1389.70 & 135.76 & 13.86 & 17.64 & 35.28 & 58.81 \\
80824 & 10.10 & 247.5755 & -12.6597 & 234.510 & -93.61 & -1184.90 & 13.91 & 17.59 & 35.18 & 58.63 \\
439 & 8.56 & 1.3346 & -37.3517 & 229.330 & 5634.07 & -2337.94 & 14.22 & 17.20 & 34.40 & 57.33 \\
15689 & 12.16 & 50.5232 & -13.2781 & 227.450 & -112.94 & -299.04 & 14.34 & 17.06 & 34.12 & 56.86 \\
3829 & 12.37 & 12.2882 & 5.3952 & 226.950 & 1233.05 & -2710.56 & 14.37 & 17.03 & 34.06 & 56.74 \\
72509 & 12.07 & 222.3862 & -26.1112 & 221.800 & -1421.60 & -203.60 & 14.70 & 16.64 & 33.28 & 55.45 \\
86162 & 9.15 & 264.1100 & 68.3422 & 220.850 & -320.47 & -1269.55 & 14.77 & 16.57 & 33.14 & 55.21 \\
85523 & 9.38 & 262.1644 & -46.8931 & 220.430 & 573.32 & -879.84 & 14.80 & 16.53 & 33.05 & 55.11 \\
\hline \hline
\end{tabular}
\label{tab:hipparcos_nearby}
\end{table}

\clearpage

\section{Star direction measurements} 
\label{app:pinhole}

In this formulation, stellar images are acquired at the spacecraft location and directions to stars are extracted through centroiding algorithms~\cite{spratling2009survey}, retrieving their position in pixel coordinates $(u,v)$. To convert image-plane measurements into line-of-sight vectors, the pinhole camera model is adopted~\cite{ma2004invitation}. Let the camera have focal length $f$ and principal point $(u_0,v_0)$. The centroid of a star $(u_\textrm{s},v_\textrm{s})$ is mapped to normalized coordinates in the camera frame as
\begin{equation}
x_c = \frac{u_\textrm{s}-u_0}{f}, \qquad y_c = \frac{v_\textrm{s}-v_0}{f}, \qquad z_c = 1,
\end{equation}
where $f$ is expressed in pixels, i.e., $f = f_{\mathrm{mm}}/p$, with $f_{\mathrm{mm}}$ the focal length in millimeters and $p$ the pixel size. The corresponding direction vector of a star in the camera frame is $\bm{r}_c = [x_c, y_c, z_c]^\top$ which, upon normalization, gives the unit LoS vector
\begin{equation}
\hat{\bm{r}}_c = \frac{\bm{r}_c}{\|\bm{r}_c\|}
\end{equation}

This vector represents the direction of the observed star relative to the camera boresight. To express this vector in the inertial frame, two sequential rotations are applied. Denoting $\bm{R}_\textrm{cb}$ the direction cosine matrix (DCM) mapping camera coordinates into the spacecraft body frame, and $\bm{R}_\textrm{bi}$ the DCM mapping from the body frame to the inertial frame, the inertial line-of-sight direction from the spacecraft to the star is given by
\begin{equation}
\hat{\bm{r}}_{\textit{i}} = \bm{R}_\textrm{bi}\,\bm{R}_\textrm{cb}\,\hat{\bm{r}}_c
\end{equation}
Here, $\bm{R}_\textrm{cb}$ is known from the camera mounting geometry and $\bm{R}_\textrm{bi}$ from the spacecraft attitude determination system.

\section{QUEST perturbation model} 
\label{appendix:quest}

Assuming small angular errors for the line-of-sight directions, the QUEST angular error model can be used, where the perturbation $\bm{\epsilon}$ acts in the tangent plane to a true line-of-sight direction $\hat{\bm{d}}$ as 
\begin{equation} \label{eq:QUEST}
\hat{\bm{d}}_{\epsilon} \approx \hat{\bm{d}} + \bm{\epsilon}
\end{equation}
where $\hat{\bm{d}}_{\epsilon}$ is the perturbed direction. Note that this model is valid for the small-angle assumption and widely used in literature~\cite{shuster1990kalman}. For large angles, analysts are required to use the multiplicative model~\cite{mortari2009multiplicative}. Assuming the perturbation vector to have zero mean and given covariance as per the QUEST model, we have 
\begin{equation}
\mathbb{E}[\bm{\epsilon}] = \bm{0} \qquad ; \qquad
\mathbb{E}[\bm{\epsilon}\bm{\epsilon}^\top] = \sigma^2 \left(\bm{I} - \hat{\bm{d}}\hat{\bm{d}}^\top\right) \\
\end{equation}

where $\textrm{E}$ is the expectation operator and $\sigma$ is the angular uncertainty of the line-of-sight direction. Note that this model is valid for small angle assumptions, where $\sigma \ll 1$.

\bibliography{sample}

\end{document}